\documentclass[aps,prl,reprint,superscriptaddress,longbibliography,floatfix]{revtex4-2}

\usepackage{amsmath,amssymb,bm}
\usepackage{graphicx}
\usepackage{hyperref}
\usepackage{physics}
\usepackage{graphicx}

\usepackage{xcolor}
\usepackage{tikz}

\newcommand{\vbsPanelTag}[1]{%
  \tikz[baseline=(vbslabel.base)]{
    \node[
      fill=white,
      fill opacity=0.7,
      text opacity=1,
      inner sep=1pt,
      outer sep=0pt
    ] (vbslabel) {\small\bfseries (#1)};
  }%
}
\newcommand{\smPanelTag}[1]{%
  \tikz[baseline=(smlabel.base)]{
    \node[
      fill=white,
      fill opacity=0.7,
      text opacity=1,
      inner sep=1pt,
      outer sep=0pt
    ] (smlabel) {\small\bfseries (#1)};
  }%
}
\newcommand{\smContourTag}{%
  \tikz[baseline=(smcontour.base)]{
    \node[
      fill=white,
      fill opacity=0.7,
      text opacity=1,
      inner sep=1pt,
      outer sep=0pt
    ] (smcontour) {\small\bfseries \(C\)};
  }%
}

\begin{document}

\title{Gauge-constrained Spinon Complexes Near Deconfined Quantum Criticality}

\author{Zhi-Yao Ning}
\email{zyning39@gmail.com}
\affiliation{Institute of Solid-State Physics, University of Tokyo, Tokyo, 277-0882, Japan}

\author{Xue-Feng Zhang}
\affiliation{Department of Physics, University of Chongqing, Chongqing, 400044, China}

\author{Naoki Kawashima}
\affiliation{Institute of Solid-State Physics, University of Tokyo, Tokyo, 277-0882, Japan}
\affiliation{Trans-scale Quantum Science Institute, The University of Tokyo, Tokyo 113-0033, Japan}

\author{Jun Takahashi}
\email{juntakahashi@issp.u-tokyo.ac.jp}
\affiliation{Institute of Solid-State Physics, University of Tokyo, Tokyo, 277-0882, Japan}

\date{\today}

\begin{abstract}
Quantum magnets provide a microscopic platform for studying confinement and gauge-constrained structures mediated by emergent gauge fields. We investigate confined spinon complexes in the columnar valence-bond-solid (VBS) phase near deconfined quantum criticality using pinned-spin defects and quantum Monte Carlo simulations. The pinned spins act as static spinon sources with controlled positions, spin projections, and VBS vorticities, enabling measurement of defect energies and direct real-space visualization of the associated VBS domain-wall strings. For matched spinon–antispinon sources, the excitation energy saturates beyond a characteristic separation as one extended dipole reorganizes into two shorter neutral dipoles, providing energetic and real-space evidence of string breaking. We further show that domain-wall connectivity is governed by lattice-scale VBS phase offsets in addition to vorticity and spin-projection neutrality. Compatible multi-pin patterns generate connected four-spinon complexes and extended domain-wall networks that retain their global connectivity under local distortions. These results establish pinned-spin defects as a controlled tool for assembling and resolving multi-spinon structures and their confining strings in a VBS phase.
\end{abstract}

\maketitle
\emph{Introduction.---}
Confinement is a central nonperturbative phenomenon in gauge theories \cite{polyakov1975compact, polyakov1977quark}. In quantum chromodynamics, color-electric flux is concentrated into tubes between separated color charges, producing a confining potential. When sufficient energy is stored in a flux tube, dynamical matter-pair creation can screen the external charges and break the string. Confinement also binds color charges into color-singlet hadrons, including conventional mesons and baryons as well as more complex multiquark states such as tetraquarks.~\cite{wilson1974confinement, bali2005observation,anber2015strings,rmp_hadron}. Recent quantum simulators have visualized the formation, dynamics, and breaking of strings connecting gauge charges.~\cite{gonzalez2025observation,cochran2025visualizing}. These developments raise a related question in strongly correlated matter: can an emergent gauge field support analogous string breaking and qualitatively distinct multi-charge complexes?

Quantum magnets near deconfined quantum critical points (DQCPs) provide a natural setting for this question. The square-lattice N\'eel--VBS transition is conventionally described by the noncompact CP\(^1\) theory of bosonic spinons coupled to an emergent \(U(1)\) gauge field, supplemented by symmetry-allowed quadrupled monopole operators that encode the microscopic compactness of the gauge field.~\cite{senthil2004deconfined,senthil2004quantum,sandvik2007evidence,wang2017deconfined}. On the VBS side, a spinon is a \(2\pi\) vortex of the VBS angle whose core carries an unpaired spin-$1/2$, forming a four-leg junction of \(\pi/2\) domain walls~\cite{sulejmanpasic2017confinement,levin2004deconfined,banerjee2014interfaces,banerjee2016finite}. Near the DQCP, the dangerously irrelevant fourfold anisotropy produces broad domain walls with strongly reduced tension.~\cite{lou2009z,shao2016quantum}. This branched string structure permits multi-spinon connectivities absent for a single thin flux tube. What remains unclear is whether multiple spinon sources can be assembled into controlled complexes and under what conditions their domain-wall connectivity can be resolved directly in real space.



Here we introduce pinned-spin defects as controlled static spinon sources in the VBS phase of the square-lattice $J$--$Q$ model near the DQCP. 
Using stochastic series expansion quantum Monte Carlo (QMC) method \cite{sandvik1999stochastic,syljuaasen2002quantum,sandvik2010computational,kaul2013bridging}, we measure defect excitation energies and reconstruct the real-space VBS-angle texture.Matched spinon–antispinon sources form confined dipoles whose extended strings reorganize into two shorter neutral objects beyond a characteristic separation, providing energetic and real-space evidence of string breaking. We further show that domain-wall connectivity depends on spin projection, vorticity, and a lattice-scale VBS phase offset. Compatible multi-pin geometries produce connected four-spinon quadrupole and defect-seeded domain-wall networks that remain visible under local perturbations. These results establish pinned defects as a controllable probe of confinement and multi-spinon organization in a VBS phase with soft strings near DQCP.

\emph{Model and Method.---}
We investigate this problem in a square lattice VBS ground state. The Hamiltonian is defined as
\begin{equation}
H=-Q\sum_{\left< ijkl \right>}{P_{ij}P_{kl}}-h\sum_{i\in 
{\mathcal{I}_\mathrm{pin}}}{\eta_i S^z_i},
\label{eq:H}
\end{equation}
where $P_{ij}=1/4-\mathbf S_i\cdot\mathbf S_j$ is the singlet projector and the summation is taken over all pairs of parallel nearest-neighbor bonds $(i,j)$ and $(k,l)$ in a plaquette of the $L\times L$ square lattice with periodic boundary conditions. We set the unit $Q=k_B=1$ and inverse temperature $\beta=L$ to effectively observe the ground state. Here $\mathcal{I}_\mathrm{pin}$ is the set of pinned sites and $\eta_i=\pm 1$ specifies the prescribed spin orientation. In the limit $h\rightarrow\infty$, the second term imposes \(S^z_i=\eta_i 1/2\). 
The ground state of the unpinned Hamiltonian lies in the columnar VBS phase, with a nearby DQCP of the $J$–$Q$ model. The $J$–$Q$ model is used here as a controlled microscopic realization of a square-lattice $Z_4$ VBS, and its proximity to the DQCP additionally produces a broad approximate-$U(1)$ regime, making the domain walls sufficiently soft to resolve their confinement and reconnection directly in real space. 
In the QMC calculation, the pinning field is treated as fixing the spin by rejecting loop updates that would flip it, which acts as a static line defect and creates a static spinon source in the VBS phase. Details of the QMC implementation are provided in Supplemental Material~\cite{SM}.

\begin{figure}[t]
    \centering
    \includegraphics[width=0.4\linewidth]{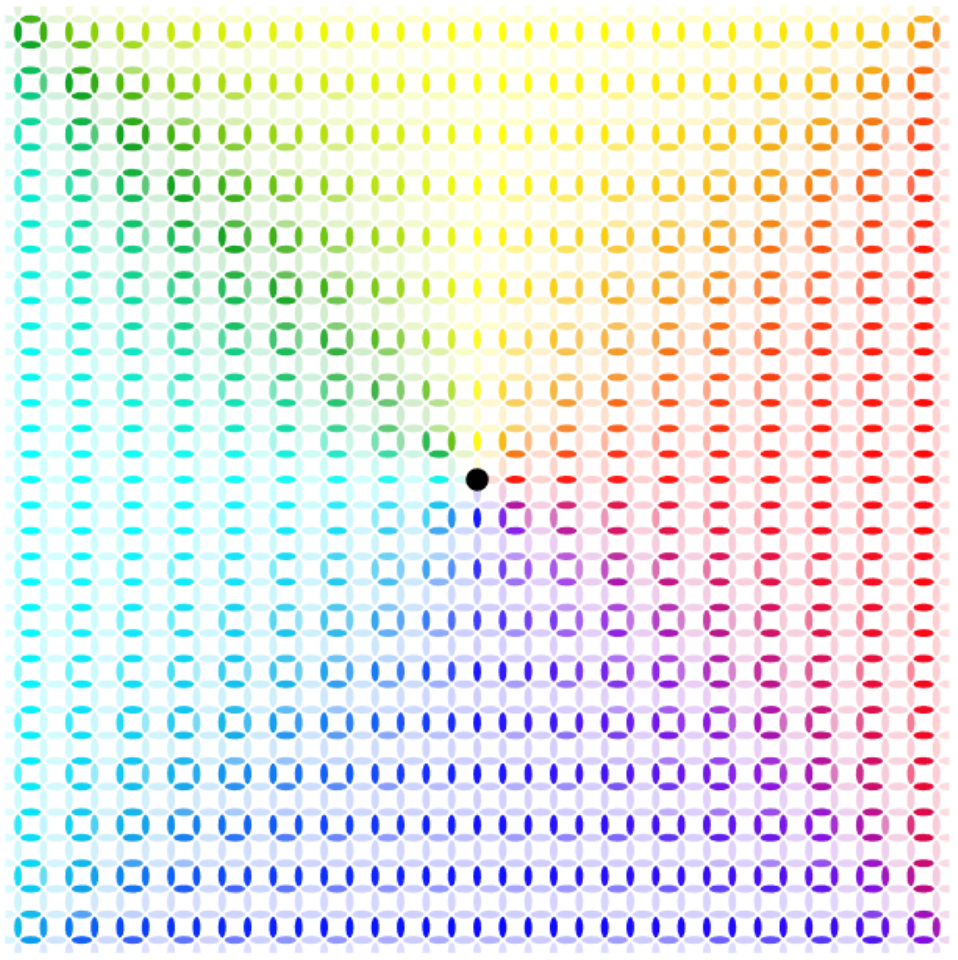}%
    \hspace{0.05\linewidth}%
    \includegraphics[width=0.42\linewidth]{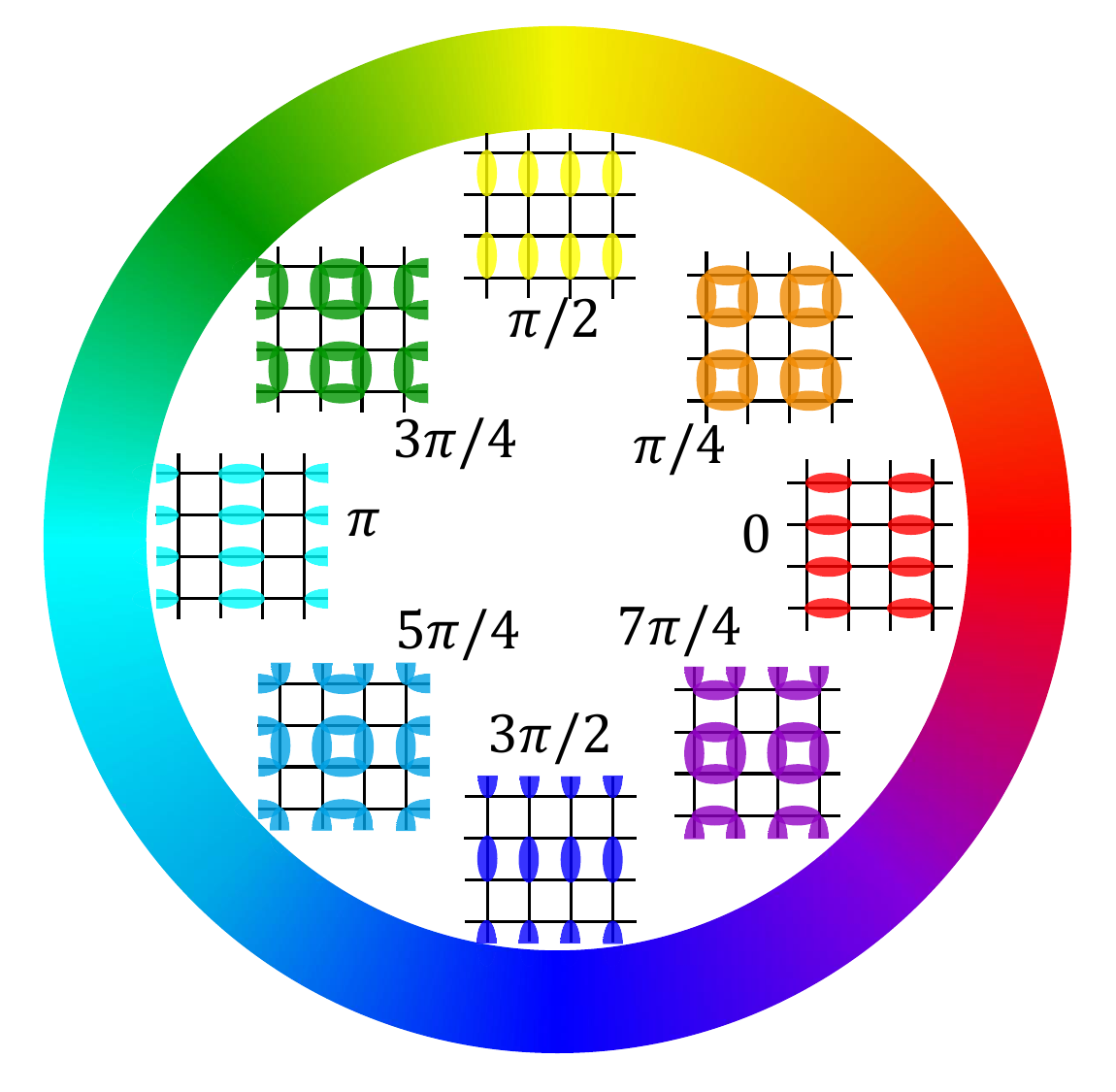}
    \caption{Left: Representative spinon texture in the VBS phase, exhibiting a $2\pi$ winding of $\phi_{\rm VBS}$. It is calculated on a odd-size square lattice with OBC, which has a natural unpaired spin as spinon. Right: Cyclic color map used to visualize the local VBS angle.}
    \label{fig:texture}
\end{figure}

To characterize and visualize the induced VBS texture, we construct a coarse-grained two-component field $\mathbf T(\mathbf b)=[T_x(\mathbf b),T_y(\mathbf b)]$ from the bond correlators following Refs.~\cite{kawashima2007ground,kaul2008imaging,zhao2020multicritical}, and define the local VBS angle on the bond \(\mathbf b\) by $\phi_{\rm VBS}(\mathbf b)=\arg[T_x(\mathbf b)+iT_y(\mathbf b)]$; the explicit coarse-graining is given in Supplemental Material~\cite{SM}. For bond-level visualization, $\phi_{\rm VBS}$ is mapped to a cyclic hue, while the magnitude of $C_{ij}=\langle S_i^zS_j^z\rangle$, which tracks the local bond energy, controls the bond transparency. The four columnar VBS states are therefore represented by four distinct colors separated by $\pi/2$, and interpolations between adjacent colors identify the $\pi/2$ domain-wall strings. A spinon is a $2\pi$ vortex of this angle field and is labeled by $(q_i,S_i^z)$, where $q_i=(2\pi)^{-1}\oint_C\nabla\phi_{\rm VBS}\cdot d\mathbf l=\pm1$ is fixed by the sublattice parity of the spinon core and $S_i^z=\pm1/2$ specifies its spin direction. A representative vortex and the color map are shown in Fig.~\ref{fig:texture}.

Within the VBS phase, the long-wavelength angular fluctuations of the VBS angle is captured by an effective sine-Gordon action~\cite{vishwanath2004quantum,lou2009z,shao2016quantum}
\begin{equation}
S_{\rm sG}=\int d\tau d^2x
\left[
\frac{K}{2}(\partial_\mu\phi_{\rm VBS})^2+\lambda_4(1-\cos 4\phi_{\rm VBS})
\right],
\label{eq:SG}
\end{equation}
where $K/2$ characterizes the stiffness and $\lambda_4$ encodes the $\mathbb{Z}_4$ anisotropy of the square lattice. The four minima of $\phi_{\rm VBS}$ encode the four columnar VBS states, and the domain walls separating them correspond to the strings. Within this normalization, the \(\pi/2\) kink has a characteristic tension $\sigma_{\rm dw}\sim \sqrt{K\lambda_4}$ and width $w_{\rm dw}\sim \sqrt{K/\lambda_4}$ as in standard sine-Gordon kink estimates~\cite{vachaspati2022kinks,manton2004topological}. Near the DQCP, the quadrupled anisotropy is dangerously irrelevant, so the effective \(Z_4\) locking is strongly suppressed on finite length scales. The same scaling implies a small effective domain-wall tension and a large wall width, providing the field-theoretic meaning of the soft VBS strings studied below~\cite{lou2009z,shao2016quantum}. Since this \(Z_4\)--\(U(1)\) crossover affects the finite-size estimate of the string tension, we focus on the \(L=48\) data when discussing the effective string tension.

For the $S^z_\mathrm{tot}=0$ sector on a periodic lattice, the total VBS vorticity must vanish because vortices of the single-valued VBS angle can only occur with zero net topological charge on the torus \cite{levin2004deconfined,senthil2004deconfined}. We write the neutrality condition of a multi-spinon configuration as:
\begin{equation}
\sum_{i\in\{\mathrm{spinons}\}}{q_i=0},\sum_{i\in\{\mathrm{spinons}\}}{S^z_i}=0,
\label{neutrality}
\end{equation}
Multiple pinned spins prescribe a subset of the spinons in such a configuration. 
If the pinned sources alone do not satisfy the required vorticity neutrality or the spin constraint of the sector considered, the low-energy texture contains additional emergent spinons that compensate the unmatched vorticity or spin projection. 
These emergent spinons play the role of dynamical matter in the gauge-theory description: unlike the externally fixed pinned sources, their positions and worldlines fluctuate within the many-body state.
Neutrality is necessary, but does not uniquely determine the domain-wall connectivity, which also depends on the lattice-scale VBS phase offsets discussed in the quantum domain-wall network section and Supplemental Material~\cite{SM}.
By varying the pinning geometry and quantum numbers, we investigate what kinds of confined bound complexes can be formed and which configurations instead require additional emergent spinons to restore the neutrality constraints.

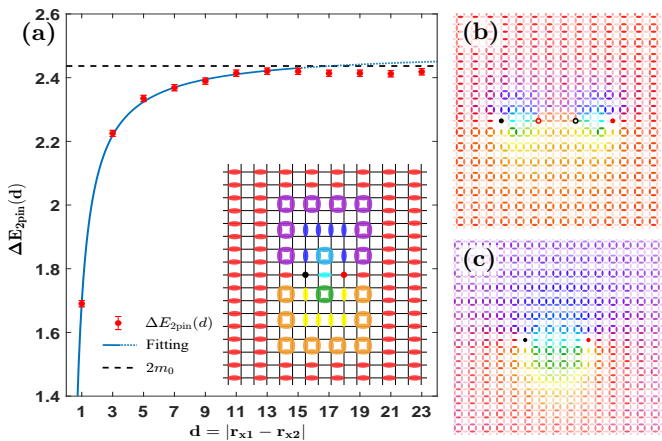
\begin{figure}[t]
    \centering
    \begin{minipage}[t]{0.66\linewidth}
        \centering
        \setlength{\unitlength}{\linewidth}
        \begin{picture}(1,1)
            \put(0,0){\includegraphics[width=\linewidth,height=\linewidth]{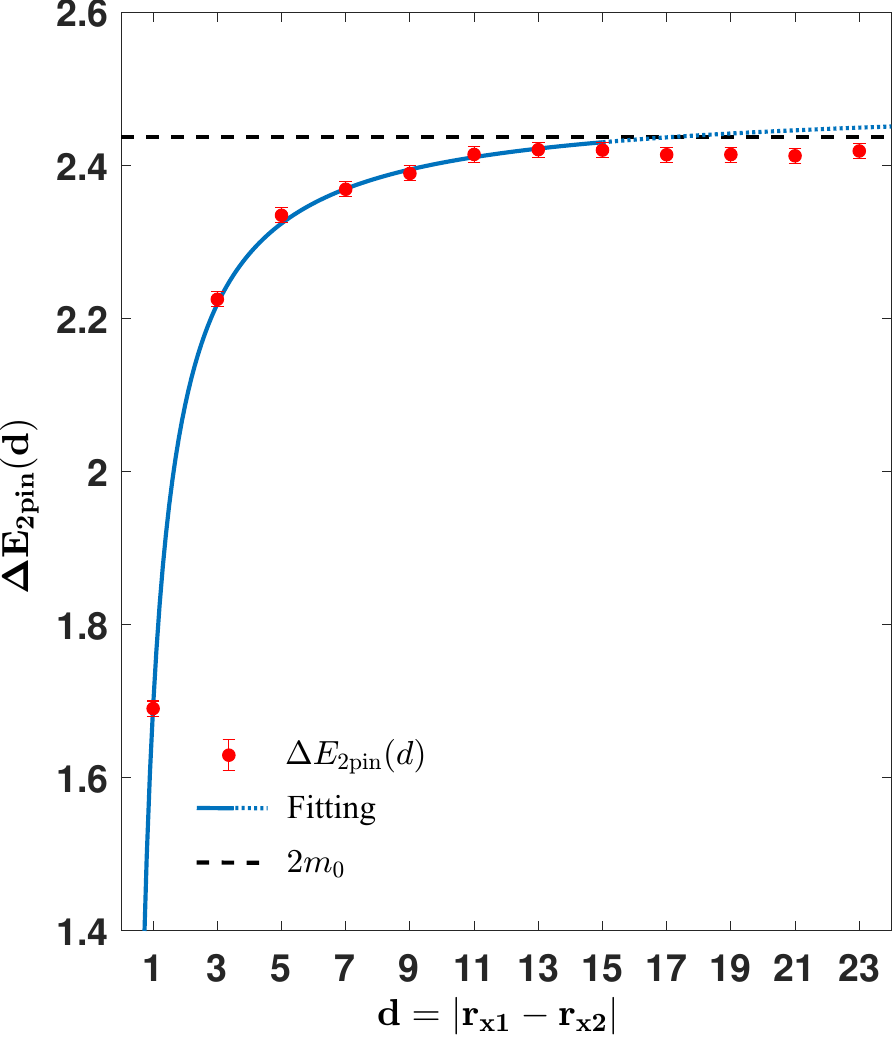}}
            \put(0.50,0.13){\includegraphics[page=1,width=0.47\linewidth]{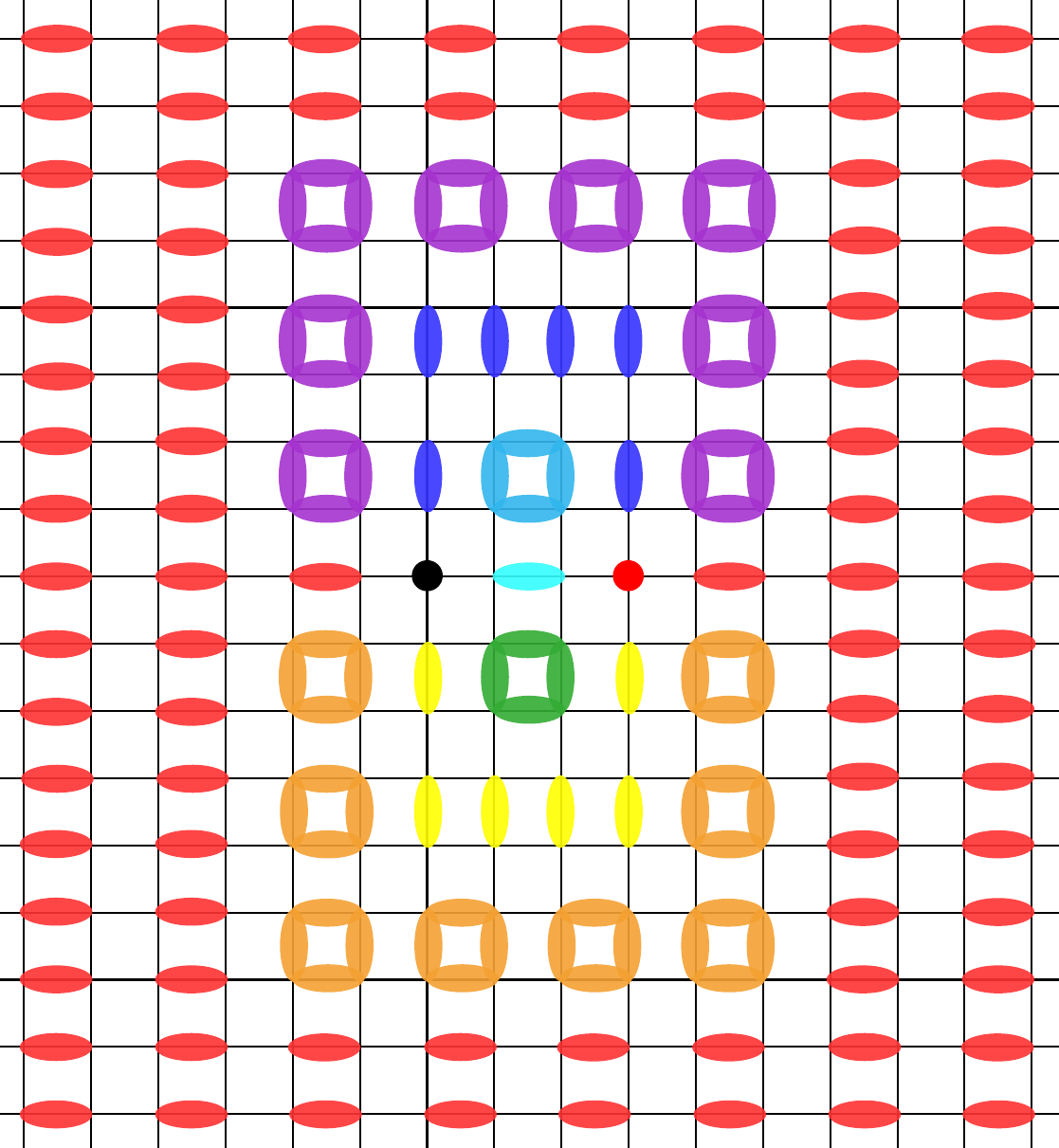}}
            \put(0.03,0.97){\makebox(0,0)[lt]{\small\bfseries (a)}}
        \end{picture}
    \end{minipage}
    \hfill
    \begin{minipage}[b]{0.319\linewidth}
        \setlength{\unitlength}{\linewidth}
        \begin{picture}(1,1.03950)
            \put(0,0){\includegraphics[width=\linewidth]{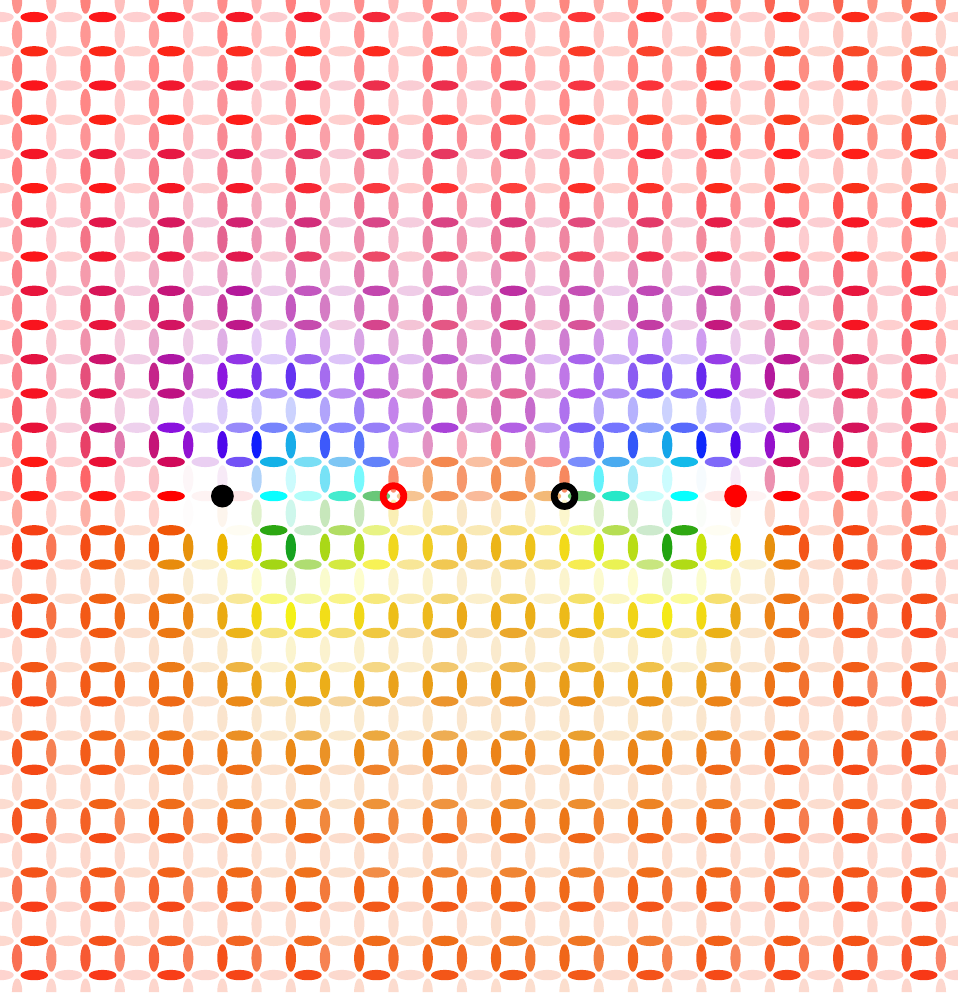}}
            \put(0.03,1.00950){\makebox(0,0)[lt]{\vbsPanelTag{b}}}
        \end{picture}\\
        \vspace{0.1cm}
        \begin{picture}(1,0.93512)
            \put(0,0){\includegraphics[width=\linewidth]{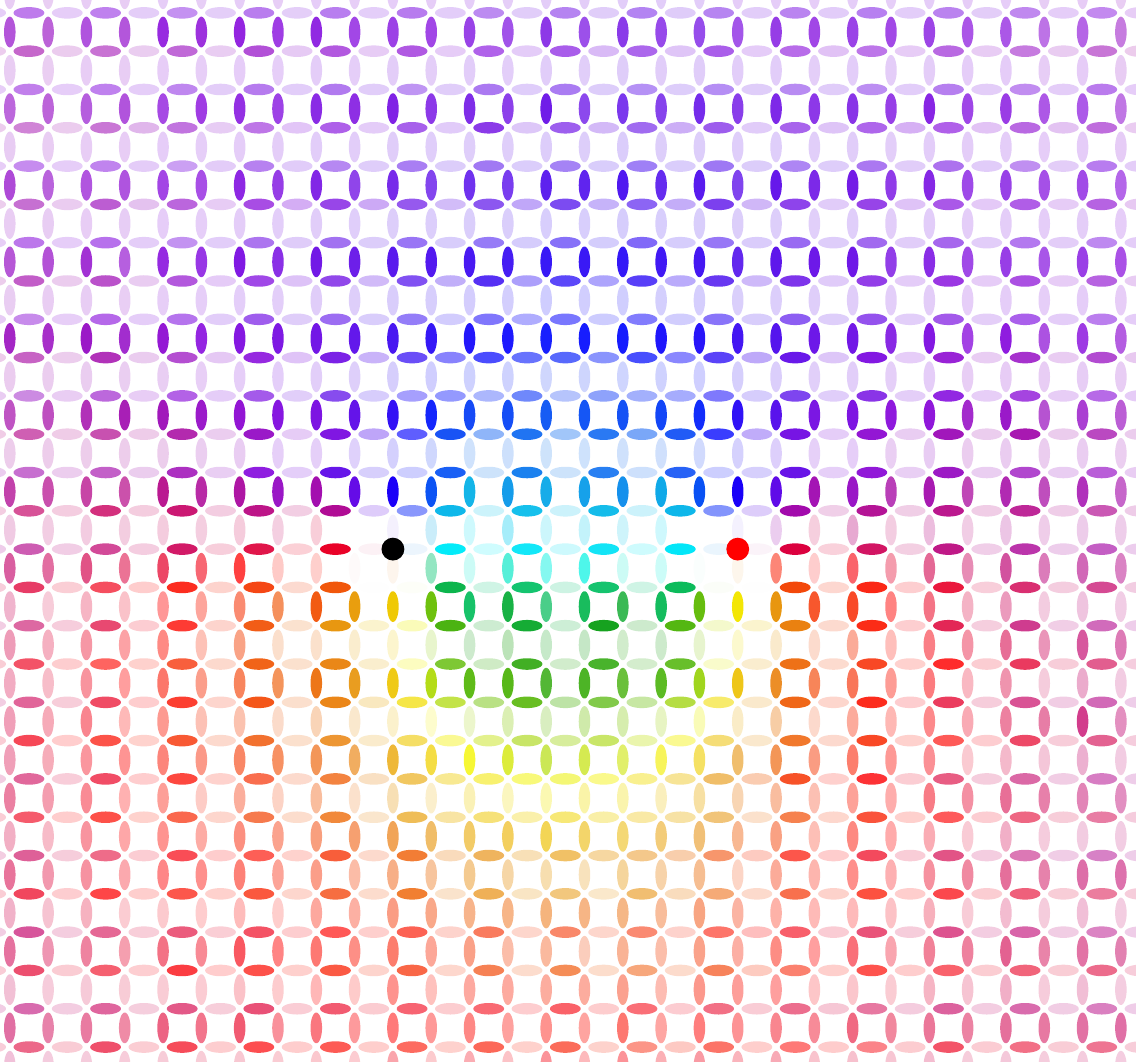}}
            \put(0.03,0.90512){\makebox(0,0)[lt]{\vbsPanelTag{c}}}
        \end{picture}
        \par\vspace{0.03\linewidth}
    \end{minipage}
    \caption{Two-pinned-spinon dipole. (a) Two-pin excitation energy \(\Delta E_{2{\rm pin}}(d)\) obtained by varying \(d\). The blue curve is a fit to the pre-breaking regime using Eq.~\eqref{eq:fit} within the fitting window \(d=1\)--\(15\), giving $\sigma\lesssim 0.001$ and \(\alpha=-0.788\pm0.0014\). Real-space VBS textures show (b) the broken and (c) the unbroken configuration. Red and black filled circles mark the pair of pinned spins with matched quantum numbers, whereas open circles mark the emergent spinons created by string breaking.}
    \label{fig:stringbreaking}
\end{figure}

\emph{Spinon dipoles.---}
We first evaluate the string breaking phenomena in the 2-pin spinon-dipole configurations. The excitation energy induced by the pinned spinons is $\Delta E_{n\mathrm{pin}}=E_{n\mathrm{pin}}-E_0$, where $E_0$ is the unpinned ground-state energy and $E_{n\mathrm{pin}}$ are the energies of n-pin configurations without the infinitely diverging magnetic field contribution (see Supplemental Material~\cite{SM} for details). We define $m_0=E_{1{\rm pin}}-E_0$ as the excitation energy of one pinned defect dressed into a neutral bound state by a dynamical (emergent) partner, and $\Delta E_{2\rm pin}(d)$ as the excitation energy of two pinned spinons spacing $d$. When the two pinned spinons hold matching quantum numbers (opposite $q$ and opposite $S^z$), they can form a single spinon-string dipole confined by four strings. The excitation energy $\Delta E_{2\mathrm{pin}}(d)$ of the spinon-string dipole grows with $d$ at short distance, reflecting the energy cost of the strings, and then saturate at large distance, as shown in Fig.~\ref{fig:stringbreaking}. 
The excitation energy increases with separation and then saturates at large \(d\). Across the same crossover, the VBS texture changes from a single extended bundle of domain walls connecting the pinned spinon–antispinon pair to two spatially separated local dipoles. Taken together, these energetic and real-space signatures are consistent with string breaking through the creation of an additional emergent spinon pair.

Since the spinon--antispinon defect is connected by several finite-width VBS domain walls, its pre-breaking energy need not follow the simple linear potential of an ideal thin flux tube. In the distance range before string breaking, we therefore use the phenomenological fitting form in analogy to the L{\"u}scher's term \cite{luscher1981symmetry,luscher2002quark}
\begin{equation}
\Delta E(d)=\Delta_\infty+\sigma d+\frac{\alpha}{d},
\label{eq:fit}
\end{equation}
where $\Delta_\infty$ approaches the broken-string plateau $2m_0$, $\sigma$ parameterizes the effective string tension. The last term $\alpha/d$ is included as a phenomenological correction that captures finite-width and internal-structure effects of the composite VBS string, including wall overlap and string fluctuations. This interpretation is motivated by the sine-Gordon description of the VBS angle, in which the approximate \(U(1)\) VBS order is weakly locked by a \(Z_4\) anisotropy \cite{vishwanath2004quantum,wang2017deconfined}. Locally, each smooth branch of the VBS domain-wall bundle can be viewed as a one-dimensional sine-Gordon kink across the wall. Since the walls have a finite width
and are repulsive to each other, 
they give 
non-additive corrections to the ideal linear string energy. 
A static wall configuration is then a kink \(\phi_{\rm VBS}\) interpolating between two adjacent \(Z_4\) minima, while being approximately uniform along tangential direction and imaginary time ~\cite{manton2004topological,vachaspati2022kinks}. When several \(\pi/2\) walls emanate from the same spinon core, their finite-width tails can overlap at intermediate separations. For walls with compatible kink orientations, such overlap is expected to generate an effective repulsive kink-kink contribution to the domain-wall energy \cite{christov2019kink, christov2019long,halcrow2023stable}. This provides a qualitative microscopic origin for non-linear corrections to the confinement potential, 
resulting in the correction in Eq.~\eqref{eq:fit} ~\cite{SM}.

We estimate the breaking distance from a winding-sector diagnostic of the sampled VBS textures. For each coarse-grained configuration, we evaluate the VBS-angle winding \(q_C=\frac{1}{2\pi}\oint_C \nabla \phi_{\rm VBS}\cdot d{\bf l}\) along several noncontractible contours $C$ crossing the region between the two pinned spinons. An unbroken string gives nonzero winding on all selected contours, whereas a broken configuration gives zero winding on at least one contour; the corresponding multi-contour criterion defines the unbroken-sector probability \(P_{\rm ub}(d)\), as detailed in Supplemental Material~\cite{SM}. 
For $L=48$, $P_{\rm ub}(d=13)>1/2$ for the longest time scale we measure, which we will use as an ad hoc definition of unbroken string. 
We therefore identify \(13<d_c\lesssim 15\) as the operational breaking distance for this system size and parameter set.

The energy fitting in Fig. \ref{fig:stringbreaking} (a) shows that the form Eq. \ref{eq:fit} explains the excess energy data well up to the string breaking distance. 
The effective linear coefficient we obtain from the fitting is small enough to be statistically indistinguishable from zero, whereas $\alpha$ remains finite in the accessible fitting window.
This near-vanishing effective tension is consistent with the finite-size \(Z_4\)–\(U(1)\) crossover near the DQCP, where the effective \(Z_4\) anisotropy is suppressed and the VBS domain walls broaden. The remaining non-linear coefficient \(\alpha\) is finite within the same fitting analysis, indicating that finite-width and internal-structure corrections dominate the pre-breaking energy over the simulated length scales. For \(d>d_c\), \(\Delta E_{\rm 2pin}(d)\) shows a small deviation from \(2m_0\), which is naturally attributed to residual interactions between the two broken dipoles.

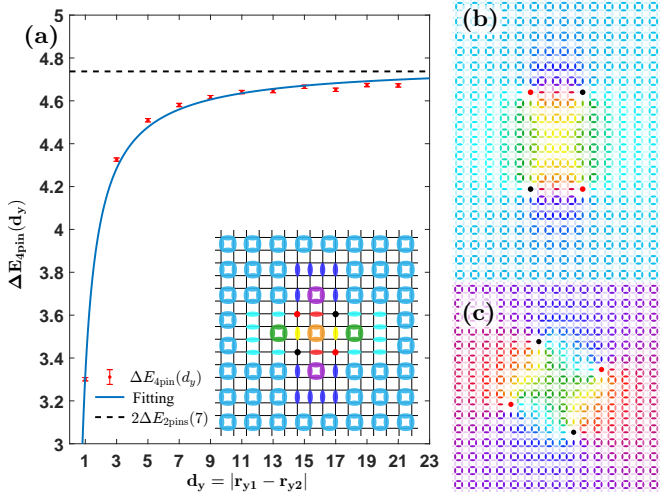
\begin{figure}[t]
    \centering
    \begin{minipage}[b]{0.66\linewidth}
        \centering
        \setlength{\unitlength}{\linewidth}
        \begin{picture}(1,1.11712)
            \put(0,0){\includegraphics[width=\linewidth]{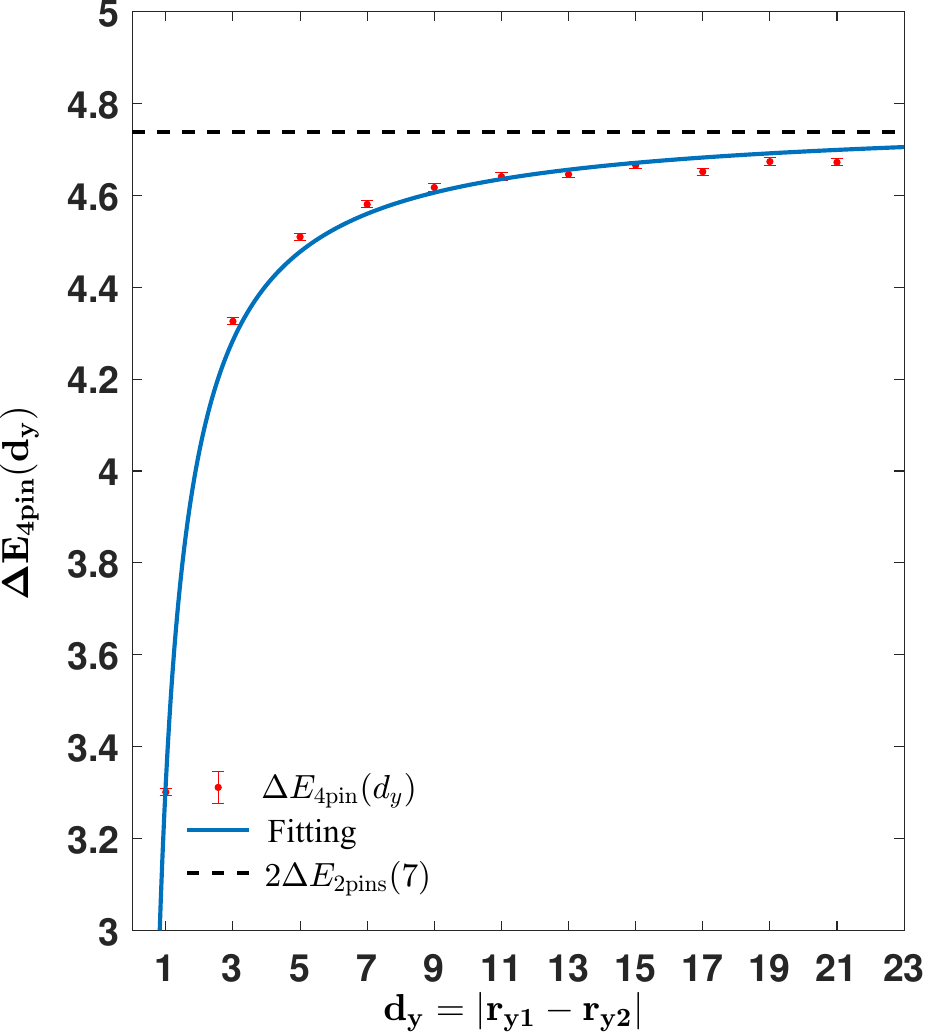}}
            \put(0.48,0.13){\includegraphics[width=0.47\linewidth]{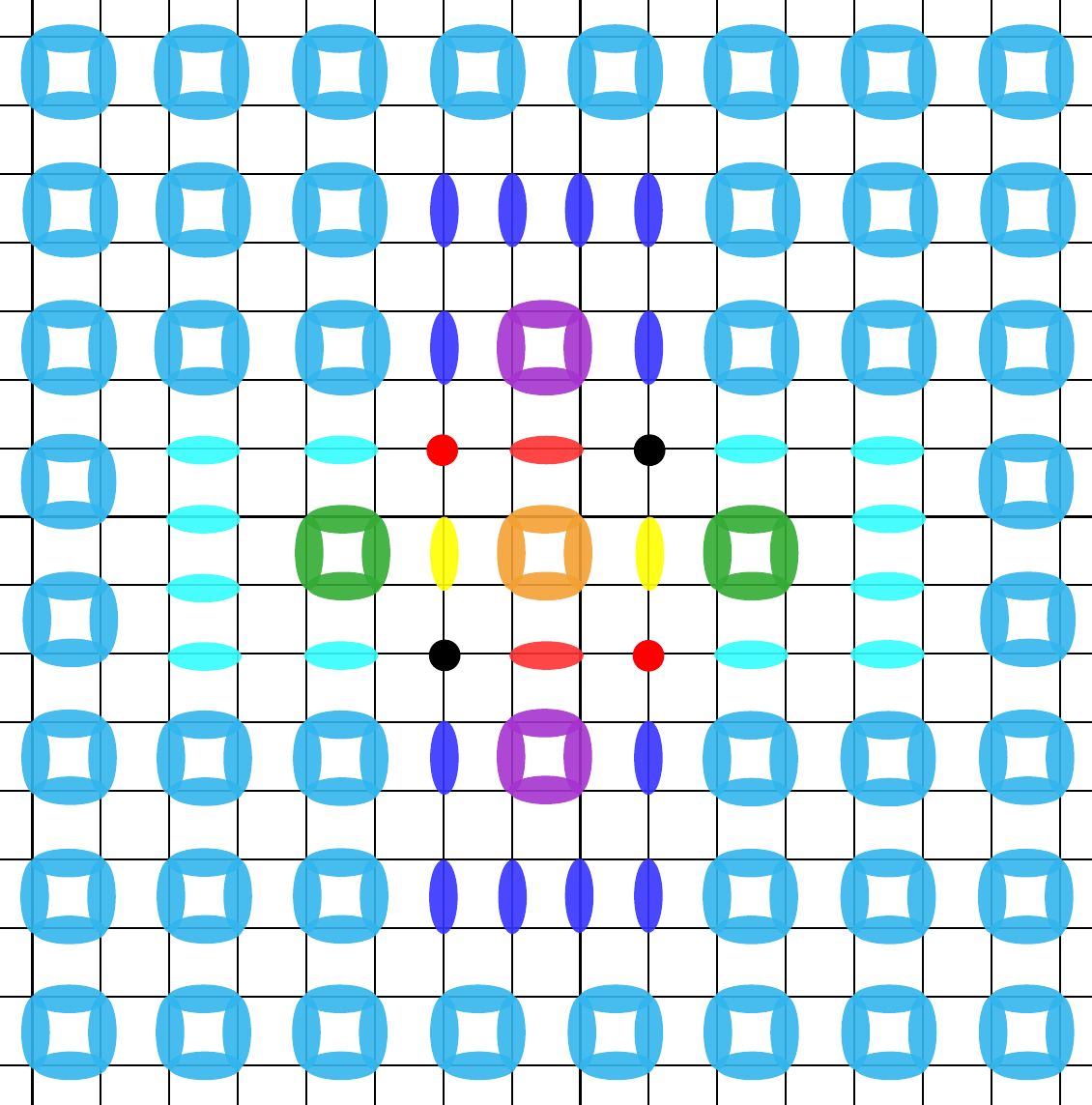}}
            \put(0.03,1.08712){\makebox(0,0)[lt]{\vbsPanelTag{a}}}
        \end{picture}
    \end{minipage}
    \hfill
    \begin{minipage}[b]{0.32\linewidth}
        \setlength{\unitlength}{\linewidth}
        \begin{picture}(1,1.35614)
            \put(0,0){\includegraphics[width=\linewidth]{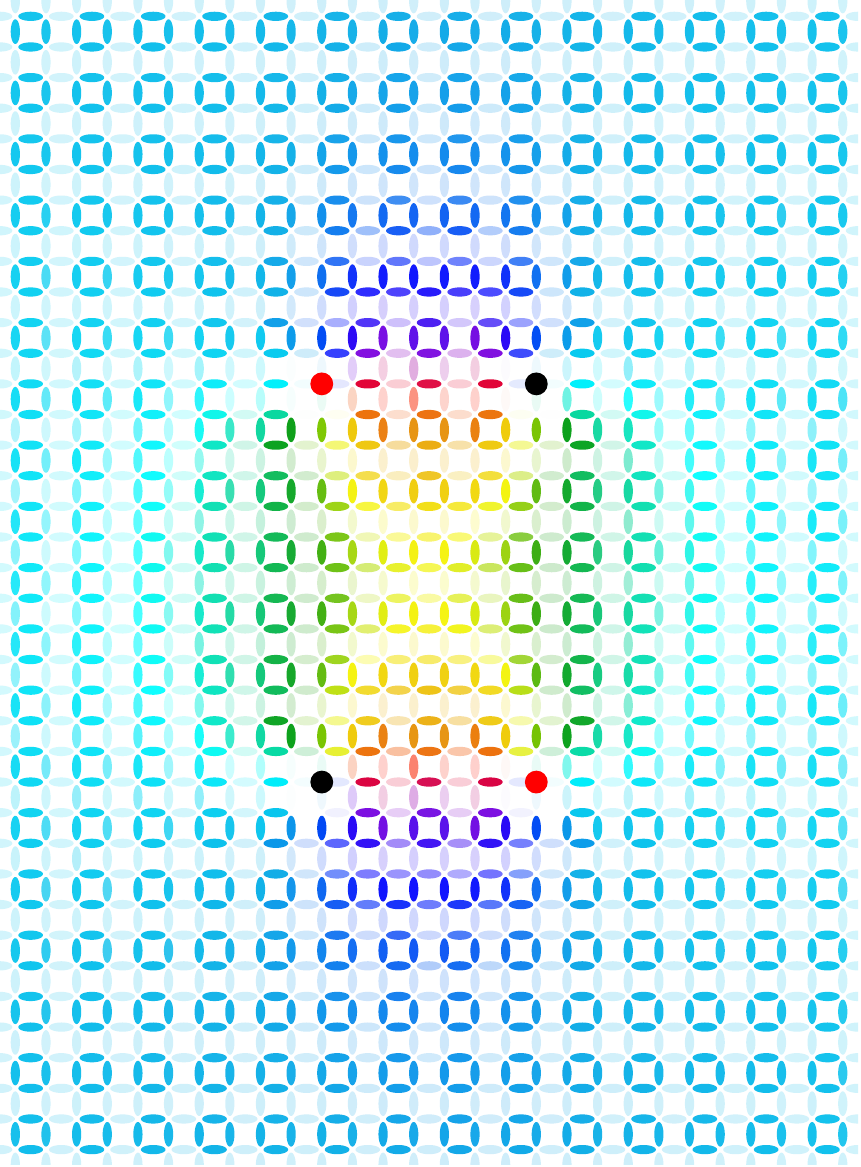}}
            \put(0.03,1.32614){\makebox(0,0)[lt]{\vbsPanelTag{b}}}
        \end{picture}\par

        \vspace{0.1cm}

        \begin{picture}(1,0.9504)
            \put(0,0){\includegraphics[width=\linewidth]{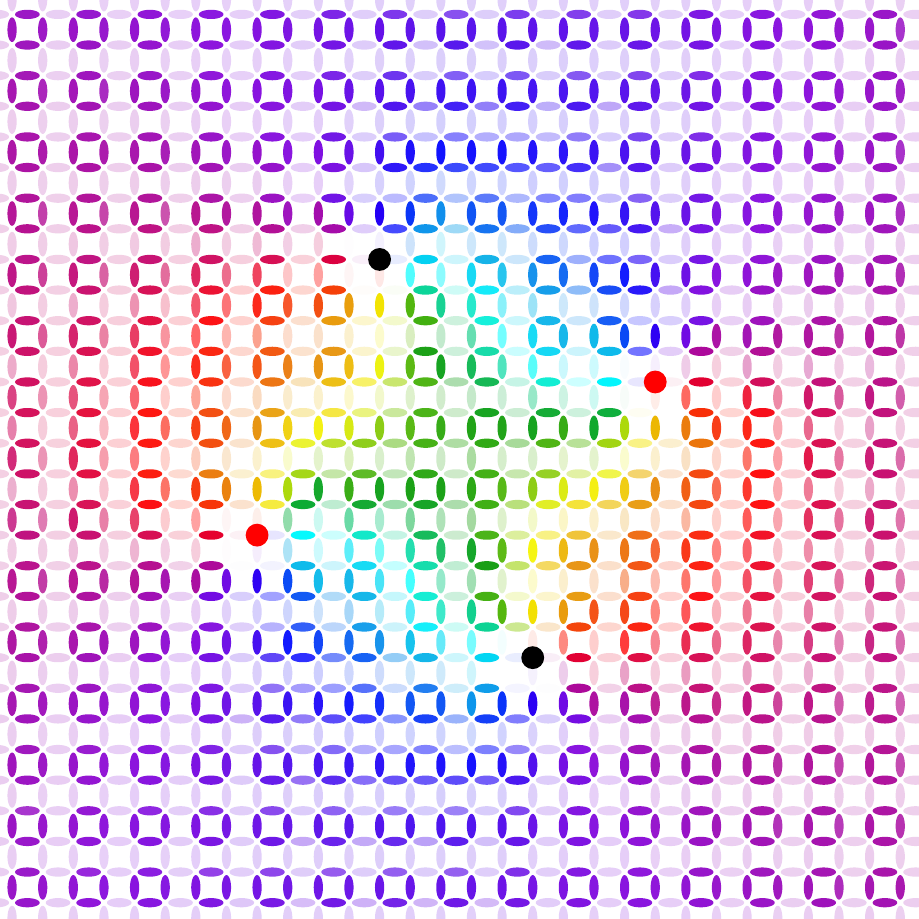}}
            \put(0.03,0.9204){\makebox(0,0)[lt]{\vbsPanelTag{c}}}
        \end{picture}
    \end{minipage}
    \caption{Four-pinned-spinon quadrupole. (a) Four-pin excitation energy \(\Delta E_{4{\rm pin}}(d_y)\) obtained by fixing \(d_x=7\) and varying \(d_y\). The blue curve is a fit to the pre-breaking form in Eq.~\eqref{eq:fit} over the window \(d_y=1\)--\(23\), giving 
    $\sigma\lesssim 0.001$
    and \(\alpha=-1.445\pm0.002\). Panels (b) and (c) show real-space VBS textures for representative four-pin configurations. The connected quadrupole remains visible when the pinned spinons are rotated with respect to the quadrupole center, as shown in (c).}
    \label{fig:4pin}
\end{figure}

\emph{Spinon quadrupoles---}
The preceding result shows that confined spinons need not organize only into isolated dipoles. We next impose four pinned spinons with alternating matched quantum numbers. As shown in Fig.~\ref{fig:4pin}, the resulting texture forms a connected four-spinon complex: a quadrupolar multi-string configuration in which neighboring spinons are connected by VBS domain-wall strings. Compared with the two-pin dipole, the domain walls in this geometry are spatially more separated, which reduces the overlap between finite-width strings and 
stabilizes the configuration. 
We compute the four-pin excitation energy \(\Delta E_{4{\rm pin}}(d_y)\) by fixing the horizontal separation \(d_x=7\) and varying the vertical separation \(d_y\). The data are fitted by the same phenomenological form as Eq.~(\ref{eq:fit}), with the fitting window and results given in the caption of Fig.~\ref{fig:4pin}. As in the two-pin case, the fitted linear coefficient is statistically indistinguishable from zero within the accessible window, while the non-linear correction remains finite. In contrast to the two-pin dipole, however, \(\Delta E_{4{\rm pin}}(d_y)\) does not show 
clear signs of string breaking with energy saturation in the 
\(L=48\) system,  
confirming 
that the four-spin complex is more stable than two dipoles. 

Notably, the four-pin geometry also reorganizes the host VBS background: while the unbroken two-pin texture retains predominantly columnar VBS order, the quadrupole locally favors plaquette-like VBS order. This behavior reflects a local selection of the VBS background by the multi-spinon geometry. We quantify this reorganization using a spatial histogram and present a detailed discussion in Supplemental Material~\cite{SM}.

Finally, we test the sensitivity of the quadrupole to local changes of the pin geometry. In the displaced configuration shown in Fig.~\ref{fig:4pin} (c), the domain-wall pattern is locally rearranged, but the connected four-spinon structure remains visible. This suggests that the quadrupole is not a fine-tuned artifact of a single lattice arrangement, but a robust finite-size texture selected by the combined spin, gauge-charge, and VBS phase-offset constraints.

\begin{figure}[t]
    \centering
    \includegraphics[width=0.46\linewidth]{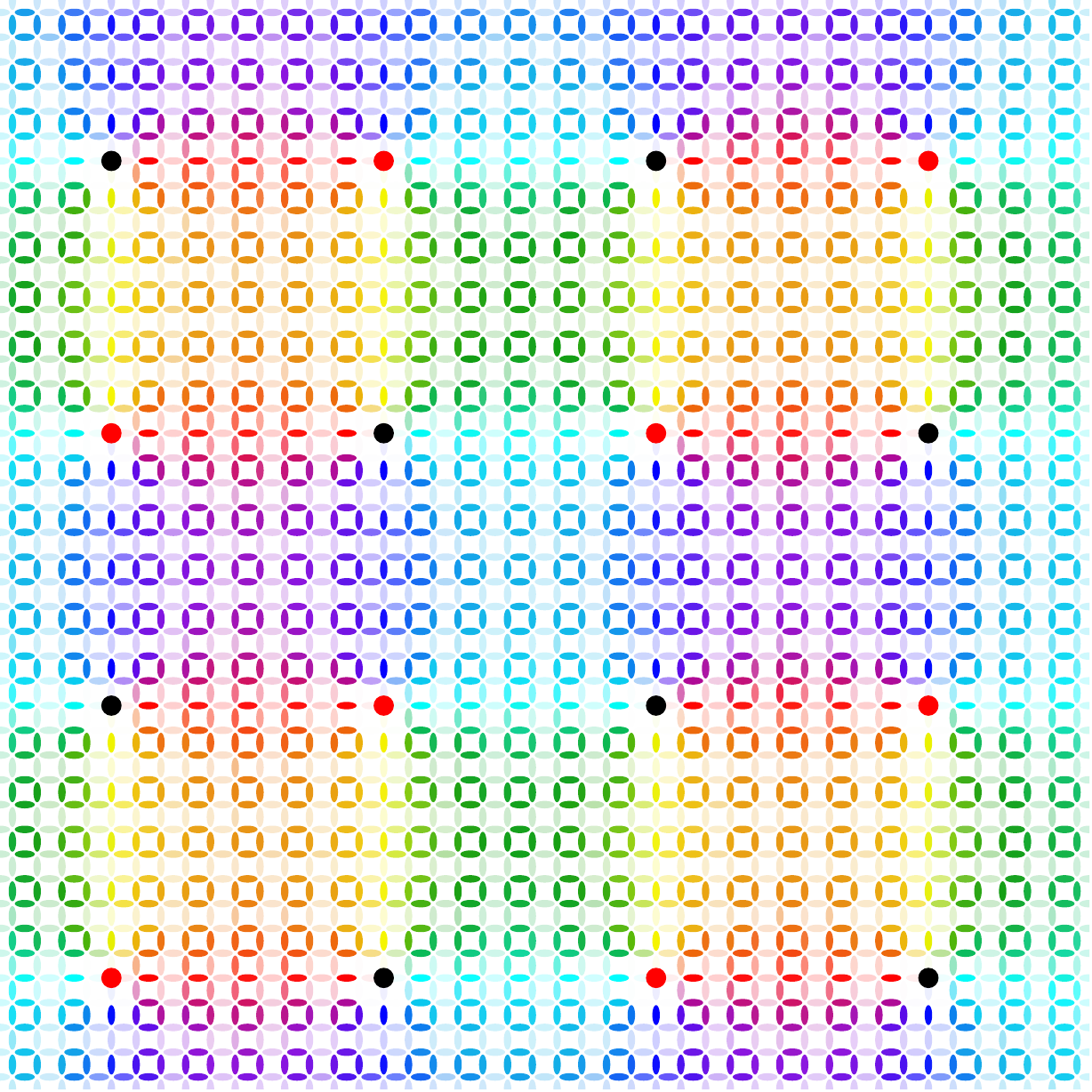}%
    \hspace{0.05\linewidth}%
    \includegraphics[width=0.463\linewidth]{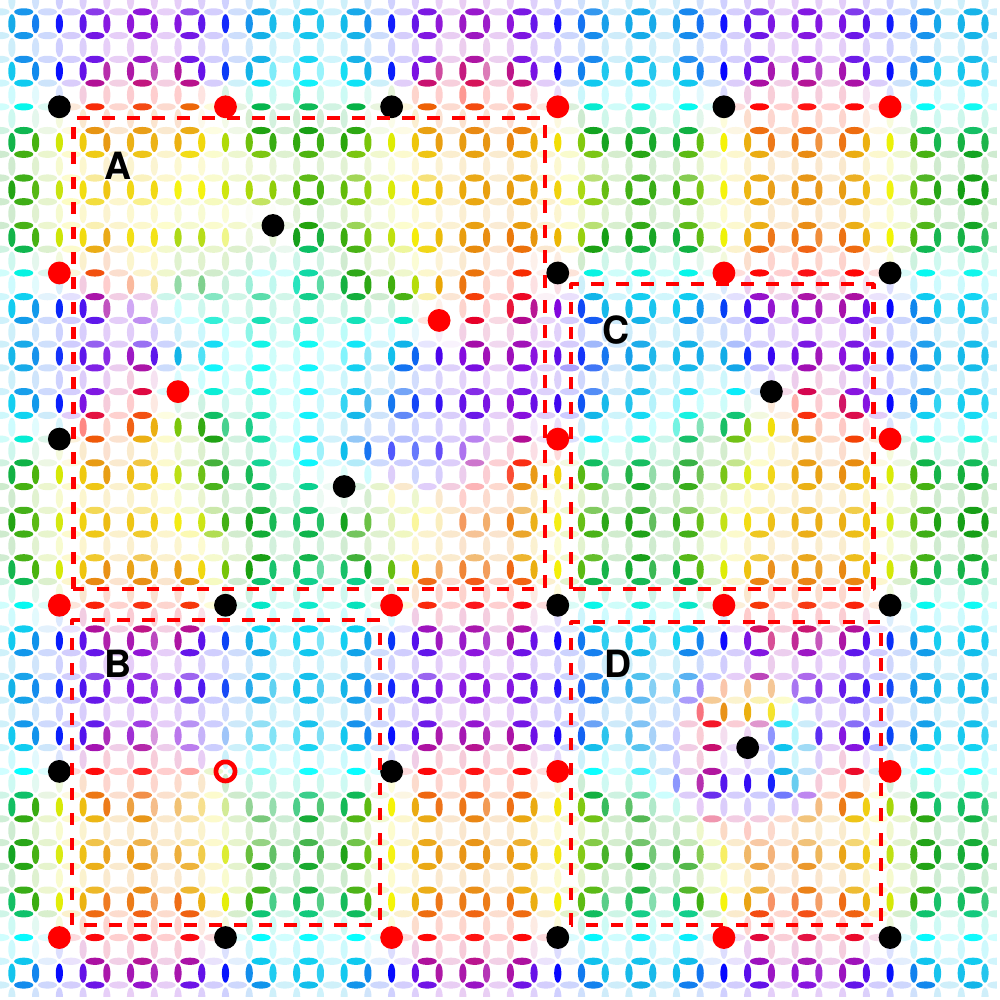}
    \caption{Domain-wall networks generated by multi-spinon pinning. (Left) regular $6\times 6$ array of pinned spinons with spacing $d=11$ produces a clean VBS domain-wall network. (Right) distorted $8\times 8$ array with smaller spacing $d=7$. The four frames mark representative perturbations, including (A) spinons displacement, (B) local dilution, (C) $\delta {\bf r}=(1,1)$ displacement, and (D) $\delta {\bf r}=(2,2)$ displacement. These perturbations deform or locally reconnect nearby strings but do not destroy the global domain-wall network.}
    \label{fig:multipin}
\end{figure}

\emph{Quantum domain wall networks.---} The four-spinon geometry also provides a controlled way to assemble extended VBS quantum domain-wall networks (QDWNs). By arranging pinned spins in a periodic and charge-compatible pattern, the imposed spinon sources act as fixed junctions of the VBS angle field. As shown in Fig.~\ref{fig:multipin}, the resulting texture is not simply a collection of independent spinon dipoles. Rather, the pinned spinons select a connected network of \(\pi/2\) domain walls whose connectivity is constrained by the vorticity and spin quantum numbers of the defect cores. Changing the spacing between pinned spinons provides a direct way to tune the network density without modifying the host Hamiltonian. The distorted configuration in the right panel of Fig.~\ref{fig:multipin} shows that local displacements, dilution, and plaquette distortions mainly rearrange nearby strings, while the global QDWN remains well defined as long as the spin, 
vorticities, and VBS phase-offset compatibility of the pinned sources is maintained. 

The apparent reversal between domain and domain wall characters follows from the VBS-angle structure. In the network texture, the enclosed regions are predominantly plaquette-like, whereas the connected walls are predominantly columnar-like. Since columnar and plaquette VBS patterns differ by a \(\pi/4\) shift of \(\phi_{\rm VBS}\), a wall interpolating between plaquette-like domains can naturally pass through columnar-like angles. Thus the columnar-looking strips should be viewed as the domain walls of a plaquette-like VBS domain network selected by the pinned-spin geometry.

\emph{Conclusion and Discussions.—}
We have used pinned-spin defects as controlled static spinon sources to probe confinement, string breaking, and multi-spinon organization in the VBS phase near a DQCP. The main conclusions are threefold. First, string breaking in this emergent setting appears as a real-space reorganization of VBS domain-wall strings: a matched pinned spinon--antispinon pair forms extended connected strings at short distance, while at larger distance the excitation energy saturates, reflecting the strings breaking into shorter neutral dipoles. Second, the allowed defect complexes are constrained not only by emergent vortex-charge neutrality, but also by net spin-projection neutrality and a lattice-scale VBS phase-offset compatibility. Representative real-space VBS patterns for mismatched two-pinned-spin configurations are presented in Supplemental Material~\cite{SM}. Third, compatible four-pin and periodic multi-pin geometries consequently generate connected quadrupole and extended domain-wall networks. The resulting geometries are shown to keep their main structures robustly under distortion. 

The pinned-defect construction in this work makes the VBS-side mechanism of string-breaking and multi-spinon complexes directly accessible in a microscopic quantum magnet that hosts VBS order such as $\mathrm{Sr}\mathrm{Cu}_2(\mathrm{BO}_3)_2$ \cite{DQCPexperiment,PhysRevLett.114.056402}. In lattice gauge theory, a string between static charges breaks when dynamically created matter screens the sources~\cite{wilson1974confinement,bali2005observation}. Here, both constituents of this description are emergent: the spinons are vortices carrying spin-\(1/2\), while the confining strings are bundles of VBS domain walls~\cite{senthil2004deconfined,senthil2004quantum,levin2004deconfined,sulejmanpasic2017confinement}. Moreover, a \(Z_4\) VBS vortex is a four-leg string junction, and its microscopic connectivity depends not only on its continuum vortex charge but also on the phase offset of the surrounding VBS pattern. This additional lattice-scale information explains why apparently equivalent pin configurations can produce different string reconnections. It also exposes a microscopic constraint that is less transparent in continuum impurity or gauge-field descriptions. Previous impurity studies in quantum magnets showed that local defects can carry spin, Berry-phase, or emergent gauge-charge quantum numbers and can nucleate VBS textures~\cite{vojta2000quantum,kolezhuk2006theory,kaul2008imaging,metlitski2008valence}. In the present lattice problem, however, a pinned spinon is also characterized by the local phase offset of the surrounding VBS pattern. Two defects with the same continuum vortex charge can therefore differ by a VBS phase offset, which can change the preferred domain-wall connectivity.


Networks of extended defects and confining strings, together with the junctions and bound complexes they support, arise in diverse settings, ranging from cosmological domain-wall webs \cite{Eto-WebOfWalls2005,2008PRD} to confining flux-tube structures in multiquark systems \cite{PentaquarkandTetraquarkStates,PentaInSU3}. 
Related structures have also been explored in condensed matter, including multi-strand confining strings in quantum dimer models and experimentally observed domain-wall networks in correlated materials \cite{banerjee2014interfaces,banerjee2016finite,park2019emergent}. More recently, Lee et al. \cite{lee2025domain} showed that a periodic network separating four domains can form a stable ground state of a continuum chiral-magnet model, where the domain-wall junctions carry fractional skyrmion charge, forming an overall skyrmion crystal. 
Our study extends this perspective, where the domain wall network is based on emergent U(1) gauge field and its junctions are the associated spinons. 
While our setup involves artificially pinned excitations, the connectivity of the domain walls is not prescribed, and instead self-consistently relaxes in the microscopic QMC simulation. 
As demonstrated in Fig.~\ref{fig:multipin}, the network is robust under source displacements, dilution, and geometric distortions. This tolerance to imperfect source configurations may be advantageous for future defect engineering attempts to realize nontrivial excitation textures in quantum materials \cite{cho2016nanoscale}. 
Although our simulations rely on the microscopic four-body interacting $Q$ terms, the underlying mechanism is not specific to it. Similar defect-induced string breaking and QDWNs can occur in other microscopic realizations of VBS order with similar spinful VBS vortices and confining domain walls.

\begin{acknowledgments}
\emph{Acknowledgments.-}
The authors thank Zi-Jian Xiong, Han Yan, Hui Shao, Zheng Yan and Anders Sandvik for valuable discussions.  This work was supported by JSPS KAKENHI Grant Numbers JP23H01092, JP25K17310, JP25H01391, JP25H01388, JP25K24845, JP25K24848 and National Natural Science Foundation of China under Grants  No.12274046, No.12547101, and Xiaomi Foundation / Xiaomi Young Talents Program.
\end{acknowledgments}

\bibliography{refs}

@misc{SM,
  note = {See Supplemental Material at [URL] for
  details of the VBS-angle construction, the statistical definition of the
  string-breaking distance, additional real-space VBS patterns, and the
  pinned-spin implementation in the stochastic series expansion.}
}

@article{rmp_hadron,
  title = {Colloquium: Hadron production in open-charm meson pairs at ${e}^{+}{e}^{\ensuremath{-}}$ colliders},
  author = {Wang, Xiongfei and Liu, Xiang and Gao, Yuanning},
  journal = {Rev. Mod. Phys.},
  volume = {98},
  issue = {2},
  pages = {021001},
  numpages = {20},
  year = {2026},
  month = {Apr},
  publisher = {American Physical Society},
  doi = {10.1103/2mrp-chly},
  url = {https://link.aps.org/doi/10.1103/2mrp-chly}
}

@article{christov2019kink,
  title={Kink-kink and kink-antikink interactions with long-range tails},
  author={Christov, Ivan C and Decker, Robert J and Demirkaya, A and Gani, Vakhid A and Kevrekidis, PG and Khare, Avinash and Saxena, Avadh},
  journal={Physical review letters},
  volume={122},
  number={17},
  pages={171601},
  year={2019},
  publisher={APS}
}

@article{christov2019long,
  title={Long-range interactions of kinks},
  author={Christov, Ivan C and Decker, Robert J and Demirkaya, A and Gani, Vakhid A and Kevrekidis, PG and Radomskiy, RV},
  journal={Physical Review D},
  volume={99},
  number={1},
  pages={016010},
  year={2019},
  publisher={APS}
}

@article{sandvik2007evidence,
  title={Evidence for Deconfined Quantum Criticality in a Two-Dimensional Heisenberg Model with Four-Spin Interactions},
  author={Sandvik, Anders W},
  journal={Physical review letters},
  volume={98},
  number={22},
  pages={227202},
  year={2007},
  publisher={APS}
}

@article{ansari2024magnetic,
  title={Magnetic effects of nonmagnetic impurities in gapped short-range resonating valence bond spin liquids},
  author={Ansari, Md Zahid and Damle, Kedar},
  journal={Physical Review Letters},
  volume={132},
  number={22},
  pages={226504},
  year={2024},
  publisher={APS}
}

@article{legg2019spin,
  title={Spin liquid mediated RKKY interaction},
  author={Legg, Henry F and Braunecker, Bernd},
  journal={Scientific Reports},
  volume={9},
  number={1},
  pages={17697},
  year={2019},
  publisher={Nature Publishing Group UK London}
}

@book{auerbach2012interacting,
  title={Interacting electrons and quantum magnetism},
  author={Auerbach, Assa},
  year={2012},
  publisher={Springer Science \& Business Media}
}

@article{wilson1974confinement,
  title={Confinement of quarks},
  author={Wilson, Kenneth G},
  journal={Physical review D},
  volume={10},
  number={8},
  pages={2445},
  year={1974},
  publisher={APS}
}

@article{kolezhuk2006theory,
  title={Theory of quantum impurities in spin liquids},
  author={Kolezhuk, Alexei and Sachdev, Subir and Biswas, Rudro R and Chen, Peiqiu},
  journal={Physical Review B—Condensed Matter and Materials Physics},
  volume={74},
  number={16},
  pages={165114},
  year={2006},
  publisher={APS}
}

@article{vojta2000quantum,
  title={Quantum impurity dynamics in two-dimensional antiferromagnets and superconductors},
  author={Vojta, Matthias and Buragohain, Chiranjeeb and Sachdev, Subir},
  journal={Physical Review B},
  volume={61},
  number={22},
  pages={15152},
  year={2000},
  publisher={APS}
}

@article{halcrow2023stable,
  title={Stable kink-kink and metastable kink-antikink solutions},
  author={Halcrow, Chris and Babaev, Egor and others},
  journal={SIGMA. Symmetry, Integrability and Geometry: Methods and Applications},
  volume={19},
  pages={034},
  year={2023},
  publisher={SIGMA. Symmetry, Integrability and Geometry: Methods and Applications}
}

@article{zhao2020multicritical,
  title={Multicritical deconfined quantum criticality and Lifshitz point of a helical valence-bond phase},
  author={Zhao, Bowen and Takahashi, Jun and Sandvik, Anders W},
  journal={Physical Review Letters},
  volume={125},
  number={25},
  pages={257204},
  year={2020},
  publisher={APS}
}

@article{cochran2025visualizing,
  title={Visualizing dynamics of charges and strings in (2+ 1) D lattice gauge theories},
  author={Cochran, Tyler A and Jobst, Bernhard and Rosenberg, Eliott and Lensky, Yuri D and Gyawali, Gaurav and Eassa, Norhan and Will, Melissa and Szasz, Aaron and Abanin, Dmitry and Acharya, Rajeev and others},
  journal={Nature},
  volume={642},
  number={8067},
  pages={315--320},
  year={2025},
  publisher={Nature Publishing Group UK London}
}

@book{vachaspati2022kinks,
  title = {Kinks and Domain Walls: An Introduction to Classical and Quantum Solitons},
  author = {Vachaspati, Tanmay},
  publisher = {Cambridge University Press},
  year = {2022},
  doi = {10.1017/9781009290456}
}

@article{anber2015strings,
  title={Strings from domain walls in supersymmetric Yang-Mills theory and adjoint QCD},
  author={Anber, Mohamed M and Poppitz, Erich and Sulejmanpa{\v{s}}i{\'c}, Tin},
  journal={Physical Review D},
  volume={92},
  number={2},
  pages={021701},
  year={2015},
  publisher={APS}
}

@book{manton2004topological,
  title={Topological solitons},
  author={Manton, Nicholas and Sutcliffe, Paul},
  year={2004},
  publisher={Cambridge University Press}
}

@article{polyakov1977quark,
  title={Quark confinement and topology of gauge theories},
  author={Polyakov, Alexander M},
  journal={Nuclear Physics B},
  volume={120},
  number={3},
  pages={429--458},
  year={1977},
  publisher={Elsevier}
}

@article{polyakov1975compact,
  title={Compact gauge fields and the infrared catastrophe},
  author={Polyakov, Alexander M},
  journal={Physics Letters B},
  volume={59},
  number={1},
  pages={82--84},
  year={1975},
  publisher={Elsevier}
}

@article{senthil2004deconfined,
  title={Deconfined quantum critical points},
  author={Senthil, Todadri and Vishwanath, Ashvin and Balents, Leon and Sachdev, Subir and Fisher, Matthew PA},
  journal={Science},
  volume={303},
  number={5663},
  pages={1490--1494},
  year={2004},
  publisher={American Association for the Advancement of Science}
}

@article{kaul2013bridging,
  title={Bridging lattice-scale physics and continuum field theory with quantum Monte Carlo simulations},
  author={Kaul, Ribhu K and Melko, Roger G and Sandvik, Anders W},
  journal={Annu. Rev. Condens. Matter Phys.},
  volume={4},
  number={1},
  pages={179--215},
  year={2013},
  publisher={Annual Reviews}
}

@article{syljuaasen2002quantum,
  title={Quantum Monte Carlo with directed loops},
  author={Sylju{\aa}sen, Olav F and Sandvik, Anders W},
  journal={Physical Review E},
  volume={66},
  number={4},
  pages={046701},
  year={2002},
  publisher={APS}
}

@article{sandvik1999stochastic,
  title={Stochastic series expansion method with operator-loop update},
  author={Sandvik, Anders W},
  journal={Physical Review B},
  volume={59},
  number={22},
  pages={R14157},
  year={1999},
  publisher={APS}
}

@article{shao2016quantum,
  title={Quantum criticality with two length scales},
  author={Shao, Hui and Guo, Wenan and Sandvik, Anders W},
  journal={Science},
  volume={352},
  number={6282},
  pages={213--216},
  year={2016},
  publisher={American Association for the Advancement of Science}
}

@article{sachdev1999quantum,
  title={Quantum phase transitions},
  author={Sachdev, Subir},
  journal={Physics world},
  volume={12},
  number={4},
  pages={33--38},
  year={1999}
}

@article{senthil2004quantum,
  title={Quantum criticality beyond the Landau-Ginzburg-Wilson paradigm},
  author={Senthil, T and Balents, Leon and Sachdev, Subir and Vishwanath, Ashvin and Fisher, Matthew PA},
  journal={Physical Review B—Condensed Matter and Materials Physics},
  volume={70},
  number={14},
  pages={144407},
  year={2004},
  publisher={APS}
}

@article{gonzalez2025observation,
  title={Observation of string breaking on a (2+ 1) D Rydberg quantum simulator},
  author={Gonz{\'a}lez-Cuadra, Daniel and Hamdan, Majd and Zache, Torsten V and Braverman, Boris and Kornja{\v{c}}a, Milan and Lukin, Alexander and Cant{\'u}, Sergio H and Liu, Fangli and Wang, Sheng-Tao and Keesling, Alexander and others},
  journal={Nature},
  pages={1--6},
  year={2025},
  publisher={Nature Publishing Group UK London}
}

@article{bali2005observation,
  title={Observation of string breaking in QCD},
  author={Bali, Gunnar S and Neff, Hartmut and Duessel, Thomas and Lippert, Thomas and Schilling, Klaus and (SESAM Collaboration)},
  journal={Physical Review D—Particles, Fields, Gravitation, and Cosmology},
  volume={71},
  number={11},
  pages={114513},
  year={2005},
  publisher={APS}
}

@article{sulejmanpasic2017confinement,
  title={Confinement in the bulk, deconfinement on the wall: infrared equivalence between compactified QCD and quantum magnets},
  author={Sulejmanpasic, Tin and Shao, Hui and Sandvik, Anders W and {\"U}nsal, Mithat},
  journal={Physical Review Letters},
  volume={119},
  number={9},
  pages={091601},
  year={2017},
  publisher={APS}
}

@article{banerjee2016finite,
  title={Finite-volume energy spectrum, fractionalized strings, and low-energy effective field theory for the quantum dimer model on the square lattice},
  author={Banerjee, D and B{\"o}gli, M and Hofmann, CP and Jiang, F-J and Widmer, P and Wiese, U-J},
  journal={Physical Review B},
  volume={94},
  number={11},
  pages={115120},
  year={2016},
  publisher={APS}
}

@article{lee2025domain,
  title={Domain wall networks as skyrmion crystals in chiral magnets},
  author={Lee, Seungho and Fujimori, Toshiaki and Nitta, Muneto and Kim, Se Kwon},
  journal={Physical Review B},
  volume={112},
  number={6},
  pages={064422},
  year={2025},
  publisher={APS}
}

@article{wang2017deconfined,
  title={Deconfined quantum critical points: symmetries and dualities},
  author={Wang, Chong and Nahum, Adam and Metlitski, Max A and Xu, Cenke and Senthil, T},
  journal={Physical Review X},
  volume={7},
  number={3},
  pages={031051},
  year={2017},
  publisher={APS}
}

@article{banerjee2014interfaces,
  title={Interfaces, strings, and a soft mode in the square lattice quantum dimer model},
  author={Banerjee, Debasish and B{\"o}gli, Michael and Hofmann, CP and Jiang, F-J and Widmer, Philippe and Wiese, U-J},
  journal={Physical Review B},
  volume={90},
  number={24},
  pages={245143},
  year={2014},
  publisher={APS}
}

@article{luscher2002quark,
  title={Quark confinement and the bosonic string},
  author={L{\"u}scher, Martin and Weisz, Peter},
  journal={Journal of High Energy Physics},
  volume={2002},
  number={07},
  pages={049--049},
  year={2002}
}

@article{luscher1981symmetry,
  title={Symmetry-breaking aspects of the roughening transition in gauge theories},
  author={L{\"u}scher, M},
  journal={Nuclear Physics B},
  volume={180},
  number={2},
  pages={317--329},
  year={1981},
  publisher={Elsevier}
}

@inproceedings{sandvik2010computational,
  title={Computational studies of quantum spin systems},
  author={Sandvik, Anders W},
  booktitle={AIP Conference Proceedings},
  volume={1297},
  number={1},
  pages={135--338},
  year={2010},
  organization={American Institute of Physics}
}

@article{levin2004deconfined,
  title={Deconfined quantum criticality and N{\'e}el order via dimer disorder},
  author={Levin, Michael and Senthil, Todadri},
  journal={Physical Review B—Condensed Matter and Materials Physics},
  volume={70},
  number={22},
  pages={220403},
  year={2004},
  publisher={APS}
}

@article{metlitski2008valence,
  title={Valence bond solid order near impurities in two-dimensional quantum antiferromagnets},
  author={Metlitski, Max A and Sachdev, Subir},
  journal={Physical Review B—Condensed Matter and Materials Physics},
  volume={77},
  number={5},
  pages={054411},
  year={2008},
  publisher={APS}
}

@article{vishwanath2004quantum,
  title={Quantum criticality and deconfinement in phase transitions between valence bond solids},
  author={Vishwanath, Ashvin and Balents, L and Senthil, T},
  journal={Physical Review B—Condensed Matter and Materials Physics},
  volume={69},
  number={22},
  pages={224416},
  year={2004},
  publisher={APS}
}

@article{kawashima2007ground,
  title={Ground states of the SU (N) Heisenberg model},
  author={Kawashima, Naoki and Tanabe, Yuta},
  journal={Physical review letters},
  volume={98},
  number={5},
  pages={057202},
  year={2007},
  publisher={APS}
}

@article{lou2009z,
  title={Z 4 to U (1) crossover of the order-parameter symmetry in a two-dimensional valence-bond solid},
  author={Lou, Jie and Sandvik, Anders W},
  journal={Physical Review B—Condensed Matter and Materials Physics},
  volume={80},
  number={21},
  pages={212406},
  year={2009},
  publisher={APS}
}

@article{kaul2008imaging,
  title={Imaging bond order near nonmagnetic impurities in square-lattice antiferromagnets},
  author={Kaul, Ribhu K and Melko, Roger G and Metlitski, Max A and Sachdev, Subir},
  journal={Physical review letters},
  volume={101},
  number={18},
  pages={187206},
  year={2008},
  publisher={APS}
}

@article{PentaInSU3,
  title = {Study for the Pentaquark Potential in SU(3) Lattice QCD},
  author = {Okiharu, Fumiko and Suganuma, Hideo and Takahashi, Toru T.},
  journal = {Phys. Rev. Lett.},
  volume = {94},
  issue = {19},
  pages = {192001},
  numpages = {4},
  year = {2005},
  month = {May},
  publisher = {American Physical Society},
  doi = {10.1103/PhysRevLett.94.192001},
  url = {https://link.aps.org/doi/10.1103/PhysRevLett.94.192001}
}

@article{PentaquarkandTetraquarkStates,
title = {Pentaquark and Tetraquark States},
journal = {Progress in Particle and Nuclear Physics},
volume = {107},
pages = {237-320},
year = {2019},
issn = {0146-6410},
doi = {https://doi.org/10.1016/j.ppnp.2019.04.003},
url = {https://www.sciencedirect.com/science/article/pii/S0146641019300304},
author = {Yan-Rui Liu and Hua-Xing Chen and Wei Chen and Xiang Liu and Shi-Lin Zhu}
}

@article{Eto-WebOfWalls2005,
  title = {Webs of domain walls in supersymmetric gauge theories},
  author = {Eto, Minoru and Isozumi, Youichi and Nitta, Muneto and Ohashi, Keisuke and Sakai, Norisuke},
  journal = {Phys. Rev. D},
  volume = {72},
  issue = {8},
  pages = {085004},
  numpages = {18},
  year = {2005},
  month = {Oct},
  publisher = {American Physical Society},
  doi = {10.1103/PhysRevD.72.085004},
  url = {https://link.aps.org/doi/10.1103/PhysRevD.72.085004}
}

@article{2008PRD,
  title = {Dynamics of domain wall networks with junctions},
  author = {Avelino, P. P. and Martins, C. J. A. P. and Menezes, J. and Menezes, R. and Oliveira, J. C. R. E.},
  journal = {Phys. Rev. D},
  volume = {78},
  issue = {10},
  pages = {103508},
  numpages = {17},
  year = {2008},
  month = {Nov},
  publisher = {American Physical Society},
  doi = {10.1103/PhysRevD.78.103508},
  url = {https://link.aps.org/doi/10.1103/PhysRevD.78.103508}
}

@article{park2019emergent,
  title={Emergent honeycomb network of topological excitations in correlated charge density wave},
  author={Park, Jae Whan and Cho, Gil Young and Lee, Jinwon and Yeom, Han Woong},
  journal={Nature Communications},
  volume={10},
  number={1},
  pages={4038},
  year={2019},
  publisher={Nature Publishing Group UK London}
}

@article{cho2016nanoscale,
  title={Nanoscale manipulation of the Mott insulating state coupled to charge order in 1 T-TaS2},
  author={Cho, Doohee and Cheon, Sangmo and Kim, Ki-Seok and Lee, Sung-Hoon and Cho, Yong-Heum and Cheong, Sang-Wook and Yeom, Han Woong},
  journal={Nature communications},
  volume={7},
  number={1},
  pages={10453},
  year={2016},
  publisher={Nature Publishing Group UK London}
}

@article{
DQCPexperiment,
author = {Yi Cui  and Lu Liu  and Huihang Lin  and Kai-Hsin Wu  and Wenshan Hong  and Xuefei Liu  and Cong Li  and Ze Hu  and Ning Xi  and Shiliang Li  and Rong Yu  and Anders W. Sandvik  and Weiqiang Yu },
title = {Proximate deconfined quantum critical point in SrCu2(BO3)2},
journal = {Science},
volume = {380},
number = {6650},
pages = {1179-1184},
year = {2023},
doi = {10.1126/science.adc9487},
URL = {https://www.science.org/doi/abs/10.1126/science.adc9487},
eprint = {https://www.science.org/doi/pdf/10.1126/science.adc9487}
}

@article{PhysRevLett.114.056402,
  title = {Real Space Imaging of Spin Polarons in Zn-Doped ${\mathrm{SrCu}}_{2}({\mathrm{BO}}_{3}{)}_{2}$},
  author = {Yoshida, M. and Kobayashi, H. and Yamauchi, I. and Takigawa, M. and Capponi, S. and Poilblanc, D. and Mila, F. and Kudo, K. and Koike, Y. and Kobayashi, N.},
  journal = {Phys. Rev. Lett.},
  volume = {114},
  issue = {5},
  pages = {056402},
  numpages = {5},
  year = {2015},
  month = {Feb},
  publisher = {American Physical Society},
  doi = {10.1103/PhysRevLett.114.056402},
  url = {https://link.aps.org/doi/10.1103/PhysRevLett.114.056402}
}


\clearpage
\onecolumngrid

\setcounter{section}{0}
\setcounter{equation}{0}
\setcounter{figure}{0}
\setcounter{table}{0}
\renewcommand{\thesection}{S\arabic{section}}
\renewcommand{\theequation}{S\arabic{equation}}
\renewcommand{\thefigure}{S\arabic{figure}}
\renewcommand{\thetable}{S\arabic{table}}

\begin{center}
{\large\bfseries Supplemental Material for\\[0.3em]
``Gauge-constrained Spinon Complexes Near Deconfined Quantum Criticality''}
\end{center}
\vspace{1em}

\setcounter{secnumdepth}{2}

\section{Scheme of treating pinned spins in SSE}
\label{app:pinned-spin-sse}
We implement the pinned spins within the standard operator-loop formulation of the stochastic series expansion (SSE)~\cite{sandvik1999stochastic,syljuaasen2002quantum,sandvik2010computational}. Writing the Hamiltonian as \(\mathcal H=-\sum_{a,b}H_{a,b}\), where \(a\) labels the operator type and \(b\) its spatial position, the partition function is expanded as
\begin{equation}
Z=\sum_{\alpha}\sum_{n=0}^{\infty}\frac{\beta^n}{n!}
\sum_{\mathcal S_n}
\left\langle\alpha\left|
\prod_{p=1}^{n}H_{a_p,b_p}
\right|\alpha\right\rangle .
\label{eq:sse-expansion-pin}
\end{equation}
In the fixed-length representation, an operator string \(\mathcal S_M\) contains \(n\) nonidentity operators and \(M-n\) identity operators, and a configuration \((\alpha,\mathcal S_M)\) has weight
\begin{equation}
W(\alpha,\mathcal S_M)=
\frac{\beta^n(M-n)!}{M!}
\prod_{p=1}^{M}
\left\langle\alpha_p\left|
H_{a_p,b_p}
\right|\alpha_{p-1}\right\rangle ,
\label{eq:sse-weight-pin}
\end{equation}
with periodicity \(\alpha_M=\alpha_0\). For the \(J\)--\(Q\) model, the nonidentity entries are diagonal or off-diagonal components of a singlet projector \(P_{ij}=1/4-\mathbf S_i\cdot\mathbf S_j\), or products of two such projectors in a \(Q\) operator.

A spin pinned along \(\mathbf N=(\sin\theta,0,\cos\theta)\) is represented by
\begin{equation}
\left|S_p(\theta)\right\rangle
=\cos\frac{\theta}{2}\left|\uparrow\right\rangle
+\sin\frac{\theta}{2}\left|\downarrow\right\rangle .
\label{eq:pinned-spin-state}
\end{equation}
For a bond joining a free spin to this pinned state, the nonzero local matrix-element magnitudes are
\begin{equation}
w_{\rm d}(\uparrow;\theta)=\frac{1}{2}\sin^2\frac{\theta}{2},\qquad
w_{\rm d}(\downarrow;\theta)=\frac{1}{2}\cos^2\frac{\theta}{2},\qquad
w_{\rm o}(\theta)=\frac{1}{2}\sin\frac{\theta}{2}\cos\frac{\theta}{2}.
\label{eq:pinned-local-weights}
\end{equation}
where \({\rm d}\) and \({\rm o}\) denote the diagonal and off-diagonal projector components, respectively. Thus, for a pinning in a general direction, insertion, removal, and loop-flip probabilities can be obtained from the detailed balance condition by including the appropriate local factors in Eq.~(\ref{eq:sse-weight-pin}). In particular, a proposed loop flip that changes several vertices adjacent to pinned spins has the Metropolis ratio
\begin{equation}
R_{\rm flip}=
\prod_{v\in\mathcal V_p}
\frac{w_v^{\rm new}(\theta)}{w_v^{\rm old}(\theta)},
\label{eq:pinned-loop-ratio}
\end{equation}
where \(\mathcal V_p\) is the set of affected pinned-spin vertices. Equivalently, projecting the singlet projector onto the pinned state produces a local boundary operator acting on a neighboring free spin,
\begin{equation}
\left\langle S_p(\theta)\right|P_{pj}\left|S_p(\theta)\right\rangle
=\frac{1}{4}
-\frac{1}{2}\left(\sin\theta\,S_j^x+\cos\theta\,S_j^z\right).
\label{eq:pinned-boundary-operator}
\end{equation}

In the present work, all pinned spins are polarized along the \(z\) axis, \(\theta=0\) or \(\pi\). We therefore use a simpler exact representation of the allowed local weights by sampling directly in the constrained configuration space
\begin{equation}
S^z_{r_p}(\tau)=s_p,\qquad
r_p\in\mathcal P,\quad 0\leq\tau<\beta ,
\label{eq:pinned-worldline-constraint}
\end{equation}
where \(s_p=\pm 1/2\) is prescribed at initialization. This constraint is equivalent, for normalized observables, to the infinite longitudinal-pinning-field limit. The field operators themselves need not be inserted explicitly into the SSE operator string; their only effect is to prevent any update from reversing a pinned world line.

The diagonal update proceeds as in the unconstrained SSE simulation. Diagonal \(J\) and \(Q\) operators are inserted or removed with the usual detailed-balance probabilities whenever their propagated-spin matrix elements are nonzero. Whenever an accepted operator acts on a pinned site, the corresponding vertex legs are recorded. During the construction of the linked-vertex representation, every loop containing one of these marked legs is identified and assigned zero flip probability. All other loops retain the standard flip probability \(1/2\). A pinned world line that contains no operator vertices is likewise excluded from the independent \(1/2\)-probability flip applied to free spins. Consequently, diagonal operators may still be sampled on bonds adjacent to a pin, and loops may fluctuate throughout the remaining system, but no accepted update changes \(S^z_{r_p}\).

This procedure preserves the detailed balance condition within the Hilbert-space sector defined by Eq.~(\ref{eq:pinned-worldline-constraint}): a loop that intersects a pinned world line would lead outside the allowed sector and is therefore rejected, whereas the relative weights and update probabilities of all admissible configurations are unchanged. The pinned sites thus appear as static line defects extending through the full imaginary-time direction, while the surrounding spins and VBS domain walls remain dynamical.

\section{Coarse-grained VBS field and visualization}
\label{app:vbs-coarse-graining}

The local VBS angle used in the main text is obtained from a site-centered,
coarse-grained field constructed from the measured bond correlators.  For each
lattice site $\mathbf r=(r_x,r_y)$, we define
\begin{align}
T_x(\mathbf r)
=&\frac{(-1)^{r_x}}{6}
\big[
C_{\mathbf r,\mathbf r+\hat{x}}
+C_{\mathbf r+\hat{y},\mathbf r+\hat{x}+\hat{y}}
+C_{\mathbf r-\hat{y},\mathbf r+\hat{x}-\hat{y}}
\nonumber\\
&\hspace{0.9cm}
-C_{\mathbf r,\mathbf r-\hat{x}}
-C_{\mathbf r+\hat{y},\mathbf r-\hat{x}+\hat{y}}
-C_{\mathbf r-\hat{y},\mathbf r-\hat{x}-\hat{y}}
\big],
\label{eq:coarse-vbs-tx}
\end{align}
where $C_{\mathbf r,\mathbf r'}=\langle
S^z_{\mathbf r}S^z_{\mathbf r'}\rangle$, and
$T_y(\mathbf r)$ is obtained from Eq.~(\ref{eq:coarse-vbs-tx}) by interchanging
$\hat{x}\leftrightarrow\hat{y}$ and $r_x\leftrightarrow r_y$.  The staggered
factor $(-1)^{r_\mu}$ removes the alternating sign associated with columnar VBS
order.  The symmetric average over the central bond and its two transverse
neighbors suppresses lattice-scale fluctuations while retaining the spatial
variation of domain walls and vortex cores.

It is convenient to combine the two components into the complex local field
\begin{equation}
\Psi_{\rm VBS}(\mathbf r)=T_x(\mathbf r)+iT_y(\mathbf r)
=\rho(\mathbf r)e^{i\phi_{\rm VBS}(\mathbf r)},
\label{eq:local-vbs-complex-field}
\end{equation}
which defines both the local amplitude $\rho=\sqrt{T_x^2+T_y^2}$ and angle
$\phi_{\rm VBS}=\arg(T_x+iT_y)$.  For a displayed bond $b=(ij)$, we first form
the endpoint average
\begin{equation}
T_\mu(b)=\frac{T_\mu(i)+T_\mu(j)}{2},\qquad \mu=x,y,
\label{eq:bond-centered-vbs-field}
\end{equation}
and evaluate $\phi_{\rm VBS}(b)$ from this bond-centered field.  This prescription
places the visualization variable at the same spatial location as the measured
bond correlator and avoids assigning a site-centered angle to an extended bond.

In the real-space plots, the cyclic hue represents $\phi_{\rm VBS}(b)$, while
the opacity is controlled by the bond strength $v=|C_{ij}|$ so that the
strongest singlet bonds remain most visible.  We do not map the nominal interval
$0\leq v\leq1$ linearly onto the full opacity range.  Instead, the visualization
routine uses the piecewise-linear map
\begin{equation}
\alpha(v)=
\begin{cases}
a_1\dfrac{v}{v_1},
&0\leq v<v_1,\\[0.6em]
a_1+(a_2-a_1)\dfrac{v-v_1}{v_2-v_1},
&v_1\leq v<v_2,\\[0.6em]
a_2+(1-a_2)\dfrac{v-v_2}{1-v_2},
&v_2\leq v\leq 1,
\end{cases}
\label{eq:bond-opacity-map}
\end{equation}
where the intial values of $v_1$ and $v_2$ 
are selected from the bond-strength histogram: $v_1$ is chosen near the lower $10\%$, so that bonds with $v\lesssim v_1$ account for approximately 10\%
of all bonds, while $v_2$ marks the upper part of the main distribution.  The
breakpoints and opacity levels $a_1,a_2$ are then adjusted slightly to maintain
clear visual contrast between weak and strong bonds.

The four columnar VBS orientations differ by $\pi/2$ in hue, and smooth color
interpolation between them resolves the domain-wall strings.  Because the
appearance is sensitive to the opacity parameters and to human color
perception, these real-space maps provide a qualitative representation of the
VBS pattern rather than a quantitative order-parameter distribution.  The
latter is extracted from the VBS-order histograms and heat maps presented in
the final two sections of this Supplemental Material.  The same coarse-grained
angle field is also used to identify the $2\pi$ winding around a spinon and in
the multi-contour string-sector diagnostic described in the following section.

\section{Statistical definition of the breaking distance $d_c$}
\label{statistical_broken}
In practice, the QMC simulation produces $N_{\rm total}$ sampling bins and we partition the $N_{\rm total}$ bins into $N_{\rm conf}=N_{\rm total}/n_{\rm ave}$ groups of $n_{\rm ave}$ bins each. Averaging within each group gives one coarse-grained configuration estimate of the VBS angle $\phi_{\rm VBS}$. For each coarse-grained configuration, we evaluate the VBS-angle winding number
\begin{equation}
q_C=\frac{1}{2\pi}\oint_C \nabla \phi_{\rm VBS}\cdot d\mathbf l ,
\end{equation}
along a closed contour $C$. We choose $C\in\mathcal C$ as a set of noncontractible contours passing through the region between the two pinned spinons, perpendicular to the line connecting them, and closed by the periodic boundary condition. Representative contours cutting through unbroken- and broken-string configurations are indicated by the vertical black dashed lines in Figs.~\ref{fig:vbs-string-variation}(c) and \ref{fig:vbs-string-variation}(d), respectively. A single contour is not sufficient to distinguish the string sector, because in a broken-string configuration one cut may still cross a local spinon dipole and give \(\left|q_{C}\right|=1\). We therefore use a multi-contour diagnostic \(Q_{\mathcal C}=\prod_{C\in\mathcal C}\left|q_{C}\right|\).
An unbroken-string configuration gives \(\left|q_{C}\right|=1\) for all selected cuts and hence \(Q_{\mathcal C}=1\). In a broken-string configuration, at least one cut passes through the broken region and gives \(q_{C}=0\), leading to \(Q_{\mathcal C}=0\). This criterion avoids misclassifying a broken configuration as unbroken when a single contour happens to intersect only one of the short dipoles.

The unbroken-configuration probability $P_{\rm ub}(d,n_{\rm ave})$ is the fraction of groups with $|Q_{\mathcal C}|=1$. Because intra-group averaging suppresses fluctuations, $P_{\rm ub}$ is positively related to the $n_{\rm ave}$ at fixed $d$: a larger $n_{\rm ave}$ effectively filters out short-lived broken fluctuations and biases the classifier toward the dominant long-time sector. The breaking distance $d_c$ is therefore extracted from the $n_{\rm ave}$-dependence of $P_{\rm ub}(d,n_{\rm ave})$ rather than from a single threshold. 

\begin{equation}
P_{\rm ub}(d,n_{\rm ave}) = \frac{\#({\text{samples with } |Q_{\mathcal C}|=1})}
{\#({\text{coarse-grained samples}})} .
\end{equation}

\begin{figure}[htp]
    \centering
    \includegraphics[width=0.4\linewidth]{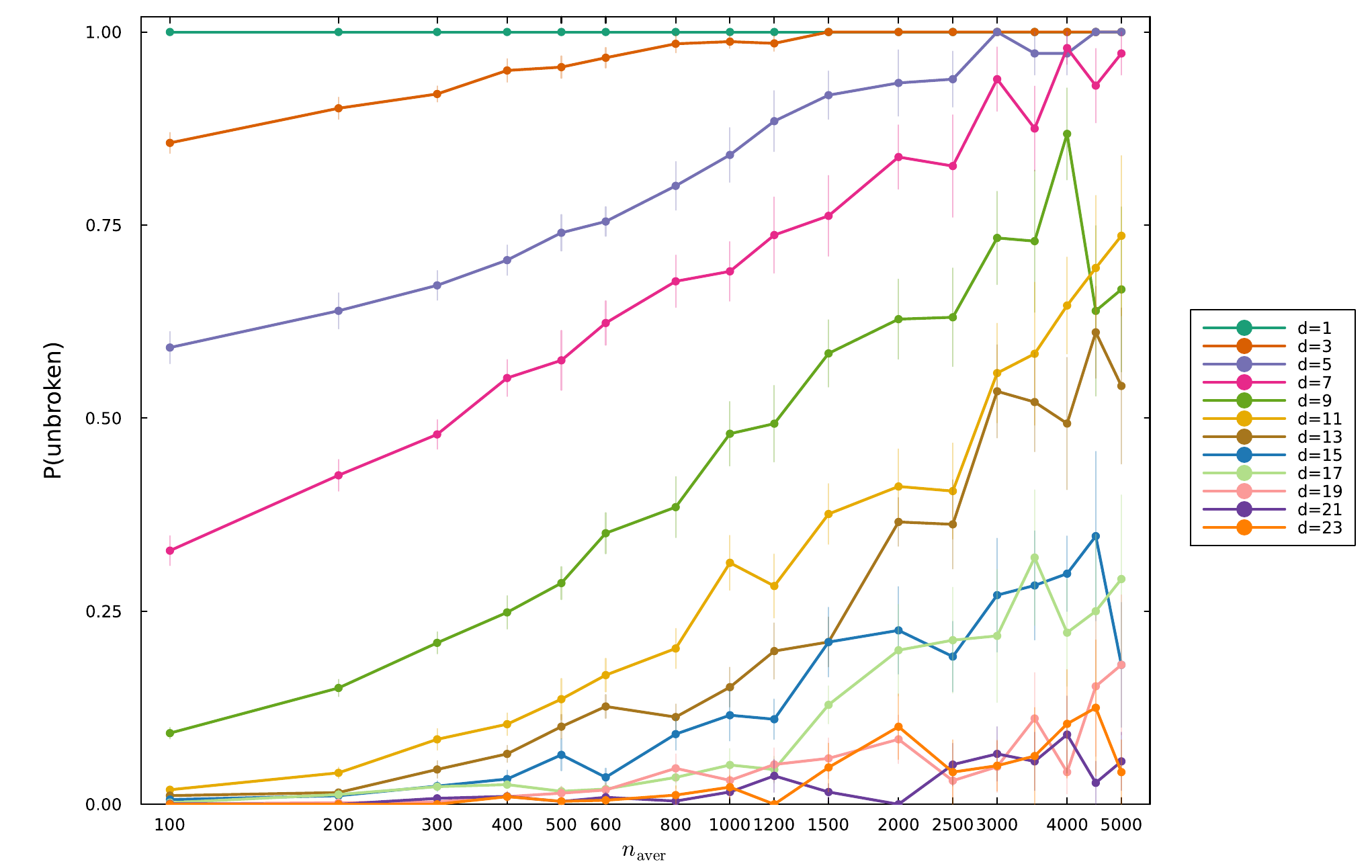}
    \caption{Unbroken-configuration probability $P_{\rm ub}(d,N)$ as a function of the number of bins $N$ averaged together for different distances $d$. }
    \label{fig:probability}
\end{figure}

The dependence on $n_{\rm ave}$ is important because the VBS strings are soft near the DQCP. Even at relatively large spinon separation, long-string configurations can occasionally appear, since the effective string tension is strongly suppressed and the energy cost of an extended string is small. Increasing $n_{\rm ave}$ filters out short-time fluctuations and emphasizes the dominant long-time texture. Therefore, if an unbroken string remains a relevant metastable sector, $P_{\rm ub}(d,n_{\rm ave})$ continues to increase with $n_{\rm ave}$. By contrast, when the system is already in the broken-string regime, the probability remains nearly independent of $n_{\rm ave}$ and stays close to the broken-sector value. In Fig.~\ref{fig:probability}, $P_{\rm ub}(d=13,n_{\rm ave})$ still grows as $n_{\rm ave}$ increases, indicating that unbroken-string configurations remain statistically relevant. For $d=15$, $P_{\rm ub}(d,n_{\rm ave})$ is nearly flat, showing that the long string is no longer a stable sector after coarse graining. We therefore identify the breaking distance as $d_c\simeq 15$ for the present system size and parameter set. In this work, all the real-space VBS patterns are plotted from coarse-grained configurations with $n_{\rm ave}=2000$ bins.

\section{Finite-width VBS domain walls and the nonlinear string energy}
\label{app:sine-gordon-string}

Here we provide a local field-theory construction underlying the interpretation
of the pre-breaking potential in the main text. The purpose is to identify the
physical understanding of corrections to an ideal linear string energy, instead of a precise quantitative description of the measured potential. 

The long-wavelength VBS angle is described by the Euclidean sine-Gordon
action~\cite{vishwanath2004quantum,wang2017deconfined}
\begin{equation}
S_{\rm sG}=\int d\tau\,d^2\mathbf r\,
\left[
\frac{K}{2}\left(\partial_\mu\phi_{\rm VBS}\right)^2
+\lambda_4\left(1-\cos 4\phi_{\rm VBS}\right)
\right].
\label{eq:sm-sg-action}
\end{equation}
The four minima \(\phi_n=n\pi/2\), with \(n=0,1,2,3\), represent the four
columnar VBS states. A domain wall between adjacent minima therefore carries
a change \(\Delta\phi_{\rm VBS}=\pi/2\). Close to the DQCP, the dangerously
irrelevant fourfold anisotropy makes the effective \(\lambda_4\) small on
finite length scales, which produces soft, broad domain walls.

\subsection{Local straight-wall solution}

\begin{figure}[htp]
    \centering
    \includegraphics[width=0.4\linewidth]{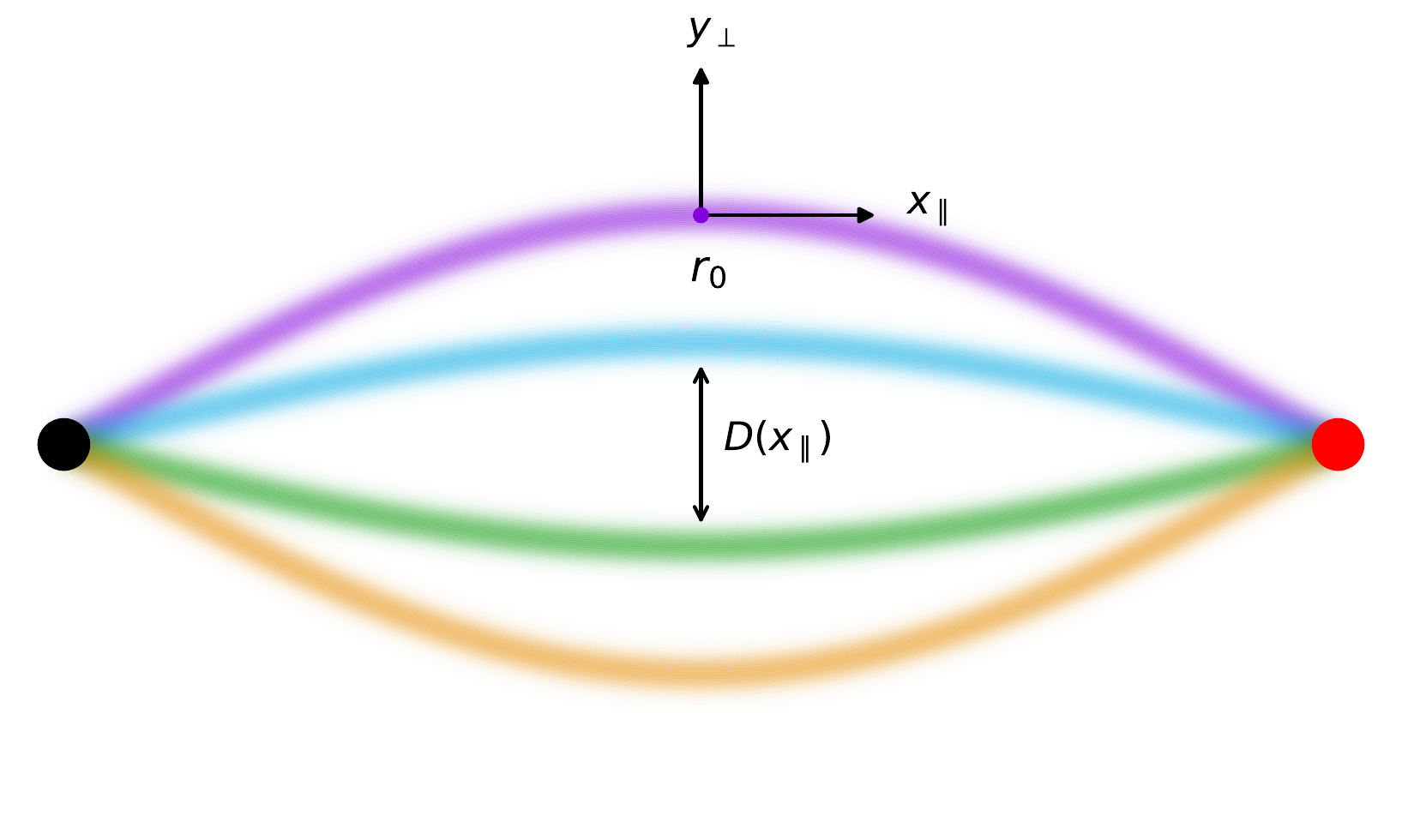}
    \caption{Local geometry of the finite-width VBS domain walls in a spinon dipole. The black and red disks denote the two pinned defects, and the four translucent ribbons represent the domain-wall branches connecting them. \(x_\parallel\) is the local tangent coordinate and \(y_\perp\) is the transverse coordinate. The separation \(D(x_\parallel)\) illustrates the local distance between neighboring branches.}
    \label{fig:sm-local-wall-geometry}
\end{figure}

Consider a smooth point \(\mathbf r_0\) on one branch of a domain wall,
sufficiently far from a spinon core or a junction. Let \(\hat{\mathbf t}\)
and \(\hat{\mathbf n}\) be the local tangent and normal vectors, respectively,
and introduce the rotated coordinates
\begin{equation}
\mathbf r=\mathbf r_0+x_\parallel\hat{\mathbf t}
+y_\perp\hat{\mathbf n}.
\label{eq:sm-local-coordinates}
\end{equation}
This local construction is illustrated in
Fig.~\ref{fig:sm-local-wall-geometry} for the composite string connecting the
two pinned defects.

If the local radius of curvature \(R_c\) is much larger than the wall width,
the branch is locally straight. A static wall is then approximately uniform
along \(x_\parallel\) and depends only on its transverse coordinate,
\begin{equation}
\phi_{\rm VBS}(x_\parallel,y_\perp,\tau)=\phi_0(y_\perp).
\label{eq:sm-straight-wall}
\end{equation}
Its energy per unit length along the wall is
\begin{equation}
\sigma[\phi_0]=\int_{-\infty}^{\infty}dy_\perp
\left[
\frac{K}{2}\left(\frac{d\phi_0}{dy_\perp}\right)^2
+\lambda_4\left(1-\cos4\phi_0\right)
\right].
\label{eq:sm-wall-energy}
\end{equation}
For a wall connecting two neighboring columnar states, we choose
\(\phi_0(-\infty)=0\) and \(\phi_0(+\infty)=\pi/2\). Minimization of
Eq.~\eqref{eq:sm-wall-energy} gives
\begin{equation}
K\frac{d^2\phi_0}{dy_\perp^2}=4\lambda_4\sin4\phi_0,
\label{eq:sm-kink-eom}
\end{equation}
together with the first integral
\begin{equation}
\frac{K}{2}\left(\frac{d\phi_0}{dy_\perp}\right)^2
=\lambda_4\left(1-\cos4\phi_0\right).
\label{eq:sm-kink-first-integral}
\end{equation}
The resulting \(\pi/2\) kink is
\begin{equation}
\phi_0(y_\perp)
=\tan^{-1}\!\left[
\exp\!\left(\frac{y_\perp-y_0}{w_{\rm dw}}\right)
\right],
\qquad
w_{\rm dw}=\frac{1}{4}\sqrt{\frac{K}{\lambda_4}},
\label{eq:sm-kink-profile}
\end{equation}
where \(y_0\) is the wall center. With the normalization in
Eq.~\eqref{eq:sm-sg-action}, its tension is
\begin{equation}
\sigma_{\rm dw}=2\sqrt{K\lambda_4}.
\label{eq:sm-wall-tension}
\end{equation}
Equations~\eqref{eq:sm-kink-profile} and \eqref{eq:sm-wall-tension} make
explicit that weak fourfold locking yields a large wall width and a small wall
tension, as in the standard sine-Gordon kink
construction~\cite{manton2004topological,vachaspati2022kinks}.

\subsection{Overlap within a composite VBS string}

A spinon is a junction of four \(\pi/2\) domain-wall branches rather than an
isolated straight wall. Nevertheless, the local kink description applies to
each smooth branch when \(w_{\rm dw}/R_c\ll1\). It fails in the spinon core and
near a junction, where the full two-dimensional vortex texture must be
retained. Away from these regions, the kink approaches its neighboring VBS
minima exponentially:
\begin{equation}
\phi_0(y_\perp)\simeq
\begin{cases}
\exp[(y_\perp-y_0)/w_{\rm dw}], & y_\perp\ll y_0,\\[2pt]
\displaystyle\frac{\pi}{2}
-\exp[-(y_\perp-y_0)/w_{\rm dw}], & y_\perp\gg y_0.
\end{cases}
\label{eq:sm-kink-tails}
\end{equation}
Consequently, two nearby branches separated by a transverse distance \(D\)
have overlapping tails. At large \(D/w_{\rm dw}\), the corresponding
interaction energy per unit wall length has the generic form
\begin{equation}
\varepsilon_{\rm int}(D)
=A\,e^{-D/w_{\rm dw}}
+O\!\left(e^{-2D/w_{\rm dw}}\right),
\label{eq:sm-overlap-density}
\end{equation}
where the sign and magnitude of \(A\) depend on the kink orientations, the
sequence of VBS minima, and the boundary conditions. Compatible neighboring
kink branches can give a repulsive leading interaction
~\cite{christov2019kink,christov2019long,halcrow2023stable}.

In a spinon dipole, neither the separation nor the curvature of the four wall
branches is constant along the composite string. If \(D(x_\parallel)\) denotes
the local branch separation, the leading overlap contribution is schematically
\begin{equation}
E_{\rm overlap}(d)
\simeq\int dx_\parallel\,
A(x_\parallel)
\exp\!\left[-\frac{D(x_\parallel)}{w_{\rm dw}}\right].
\label{eq:sm-overlap-energy}
\end{equation}
Endpoint curvature, deformation of the kink profiles, and fluctuations of the
whole bundle provide further non-additive terms. Before string breaking, the
defect energy can therefore be organized as
\begin{equation}
\Delta E(d)
=E_{\rm core}
+\sigma_{\rm eff}d
+E_{\rm overlap}(d)
+E_{\rm curv}(d)
+E_{\rm fluc}(d)+\cdots .
\label{eq:sm-energy-decomposition}
\end{equation}
This decomposition explains why a bundle of broad, interacting VBS domain
walls need not display a purely linear potential over the finite separations
accessible in the simulation.

The individual terms in Eq.~\eqref{eq:sm-energy-decomposition} cannot be
separated reliably within the pre-breaking fitting window. We therefore
represent their combined smooth distance dependence by the compact
phenomenological form used in the main text,
\begin{equation}
\Delta E(d)=\Delta_\infty+\sigma d+\frac{\alpha}{d}.
\label{eq:sm-fit-form}
\end{equation}
Here \(\Delta_\infty\) connects to the broken-string plateau \(2m_0\), while
\(\sigma\) is an effective linear coefficient. The coefficient \(\alpha\)
absorbs finite-width, overlap, curvature, and fluctuation effects over the
available range of \(d\).

\section{Real-space VBS pattern variation}
\label{VBS-variation}
\begin{figure}[htp]
    \centering
    \setlength{\unitlength}{0.24\linewidth}
    \begin{picture}(1,1)
        \put(0,0){\includegraphics[width=\unitlength]{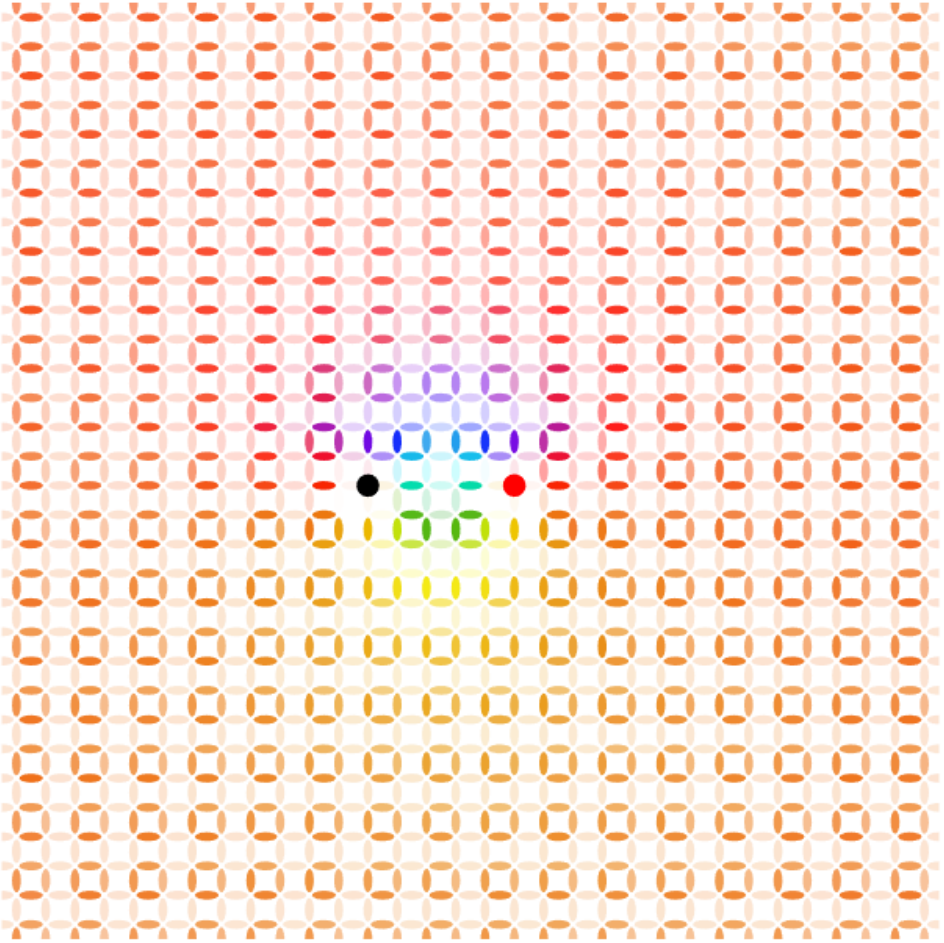}}
        \put(0.03,0.97){\makebox(0,0)[lt]{\smPanelTag{a}}}
    \end{picture}%
    \hfill
    \setlength{\unitlength}{0.24\linewidth}
    \begin{picture}(1.06938,1)
        \put(0,0){\includegraphics[height=\unitlength]{figs/P2S_L48_string_d9_unbroken.pdf}}
        \put(0.03,0.97){\makebox(0,0)[lt]{\smPanelTag{b}}}
    \end{picture}%
    \hfill
    \setlength{\unitlength}{0.225\linewidth}
    \begin{picture}(1,1.061)
        \put(0,0){\includegraphics[width=\unitlength]{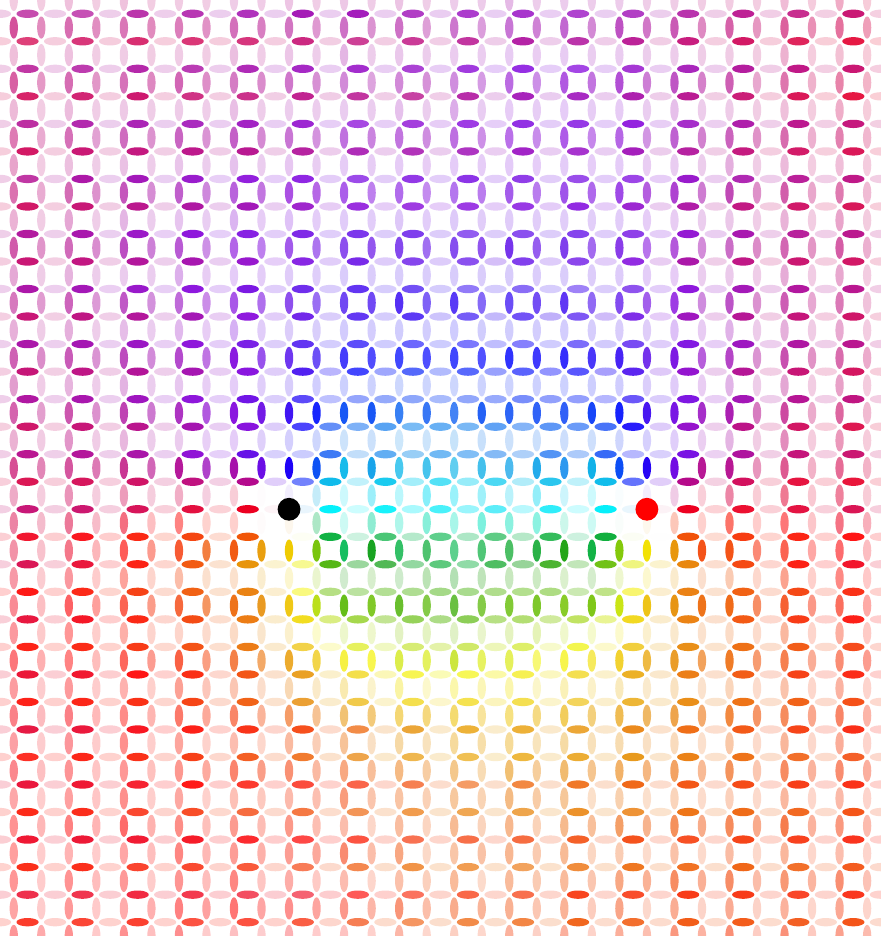}}
        {\linethickness{0.8pt}\multiput(0.5,0)(0,0.03996){27}{\line(0,1){0.022}}}
        \put(0.53,1.00){\makebox(0,0)[bl]{\smContourTag}}
        \put(0.03,1.031){\makebox(0,0)[lt]{\smPanelTag{c}}}
    \end{picture}%
    \hfill
    \setlength{\unitlength}{0.24\linewidth}
    \begin{picture}(1,1)
        \put(0,0){\includegraphics[width=\unitlength]{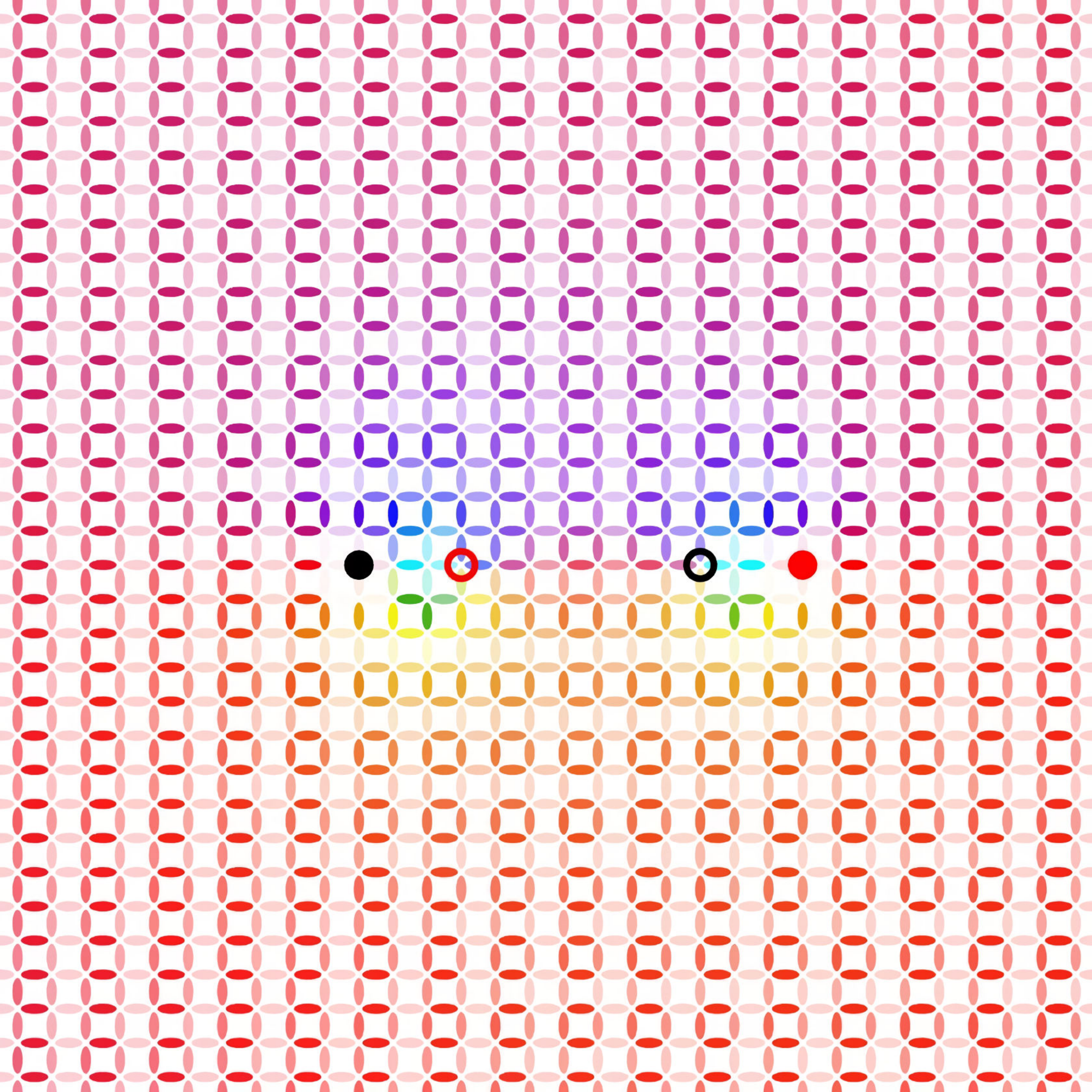}}
        {\linethickness{0.8pt}\multiput(0.5,0)(0,0.03912){26}{\line(0,1){0.022}}}
        \put(0.53,0.94){\makebox(0,0)[bl]{\smContourTag}}
        \put(0.03,0.97){\makebox(0,0)[lt]{\smPanelTag{d}}}
    \end{picture}
    \caption{Evolution and breaking of the real-space VBS pattern for a
    spinon--string dipole in the $L=48$ system.  From left to right:
    (a)~$d=5$, unbroken; (b)~$d=9$, unbroken; (c)~$d=13$, unbroken; and
    (d)~$d=13$, broken.  Increasing the pinned-spinon separation stretches the
    connected VBS domain-wall bundle.  Near the breaking crossover, illustrated
    here at $d=13$, both unbroken and broken configurations are sampled at the
    same separation. The vertical black dashed lines in (c) and (d) mark
    representative noncontractible contours $C$, closed through the periodic
    boundary condition.}
    \label{fig:vbs-string-variation}
\end{figure}

Figure~\ref{fig:vbs-string-variation} compares the real-space VBS patterns of
spinon--string dipoles at three pinned-spinon separations.  As the separation
increases from $d=5$ to $d=9$ and $d=13$, the connected domain-wall bundle is
progressively stretched between the two pinned spinons.  The analysis in the
Supplemental Material section on the
\hyperref[statistical_broken]{statistical definition of the breaking distance
\(d_c\)} shows that broken and unbroken string configurations can both occur at
the same separation near the crossover.
Accordingly, for $d=13$ we display representative configurations from both
sectors: one retains the extended connection, whereas the other has broken
into two shorter spinon--string dipoles.

\begin{figure}[htp]
    \centering
    \setlength{\unitlength}{0.24\linewidth}
    \begin{picture}(1,1.426)
        \put(0,0){\includegraphics[width=\unitlength]{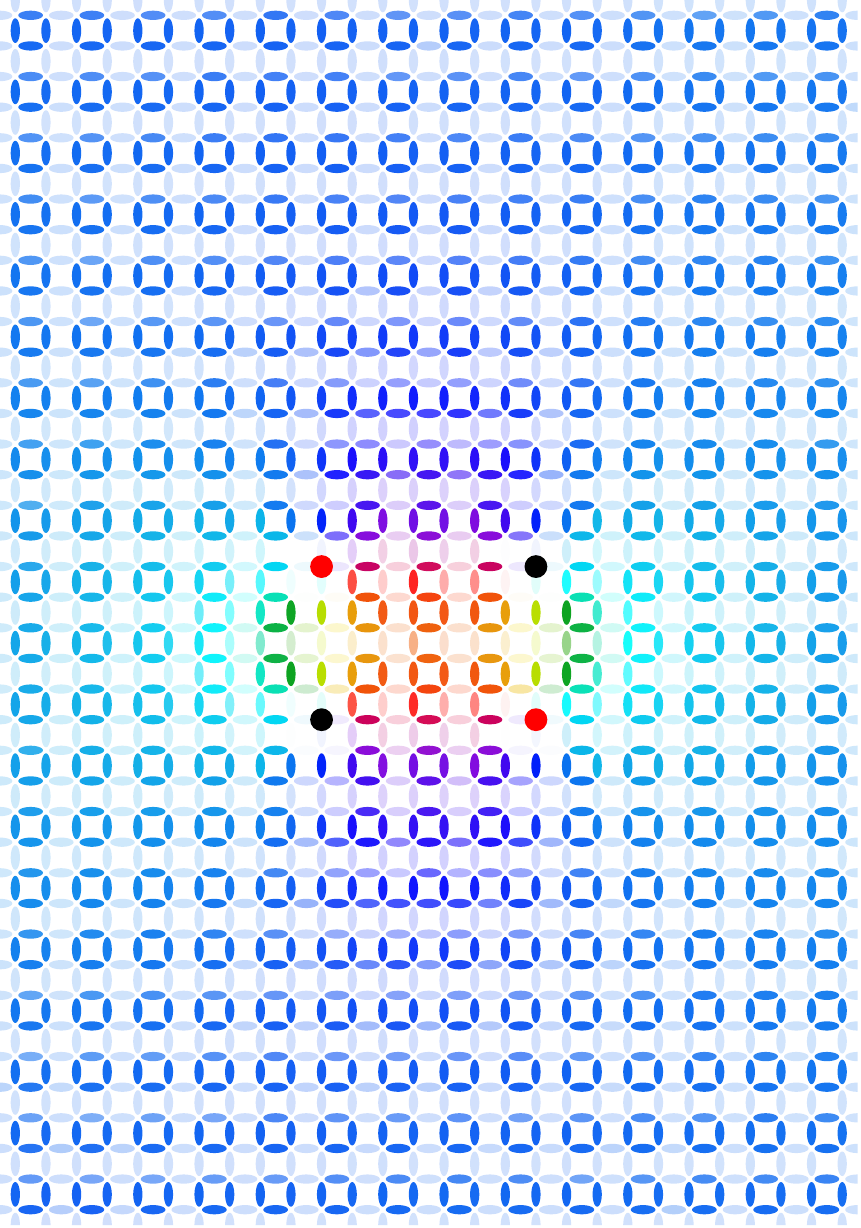}}
        \put(0.03,1.396){\makebox(0,0)[lt]{\smPanelTag{a}}}
    \end{picture}%
    \hfill
    \begin{picture}(1,1.426)
        \put(0,0){\includegraphics[width=\unitlength]{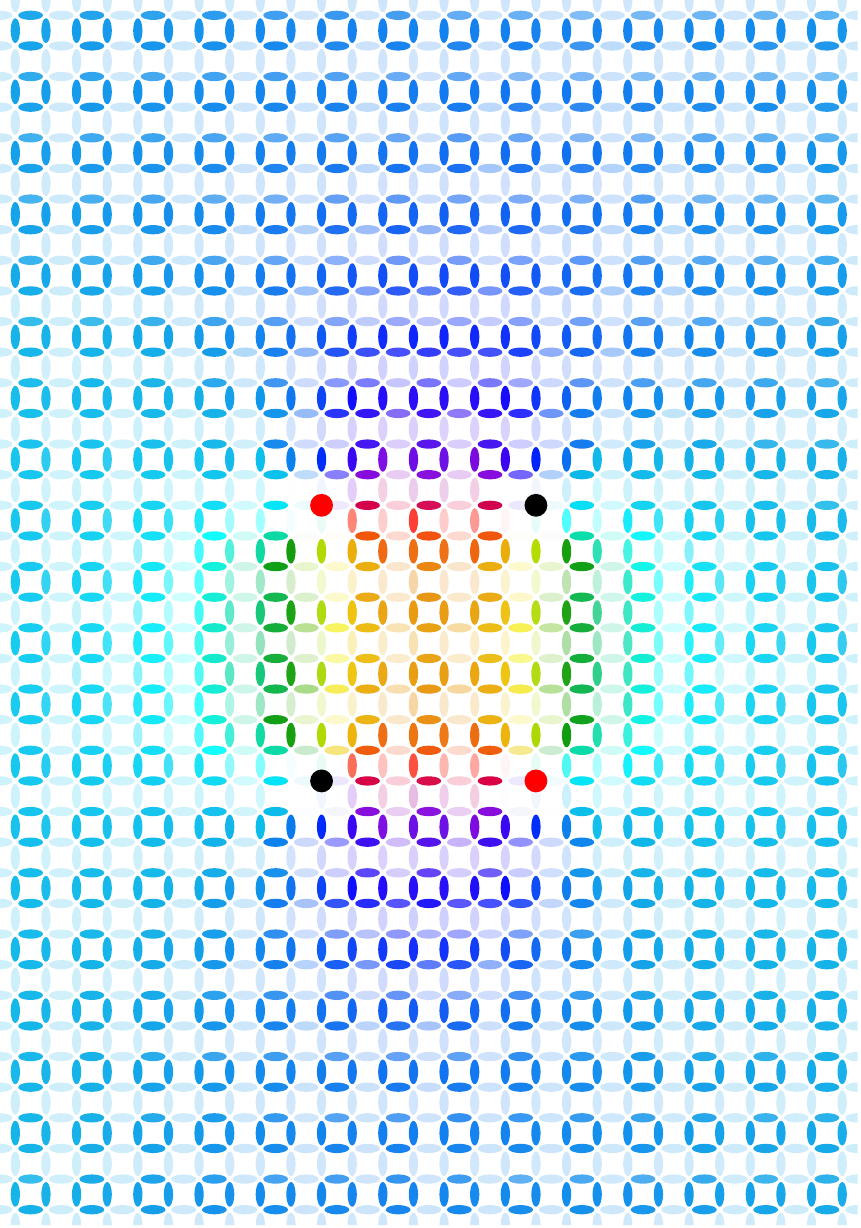}}
        \put(0.03,1.396){\makebox(0,0)[lt]{\smPanelTag{b}}}
    \end{picture}%
    \hfill
    \begin{picture}(1.05162,1.426)
        \put(0,0){\includegraphics[height=1.426\unitlength]{figs/P4S_L48_rectangle_dx3_dy6.pdf}}
        \put(0.03,1.396){\makebox(0,0)[lt]{\smPanelTag{c}}}
    \end{picture}%
    \hfill
    \begin{picture}(1,1.426)
        \put(0,0){\includegraphics[width=\unitlength]{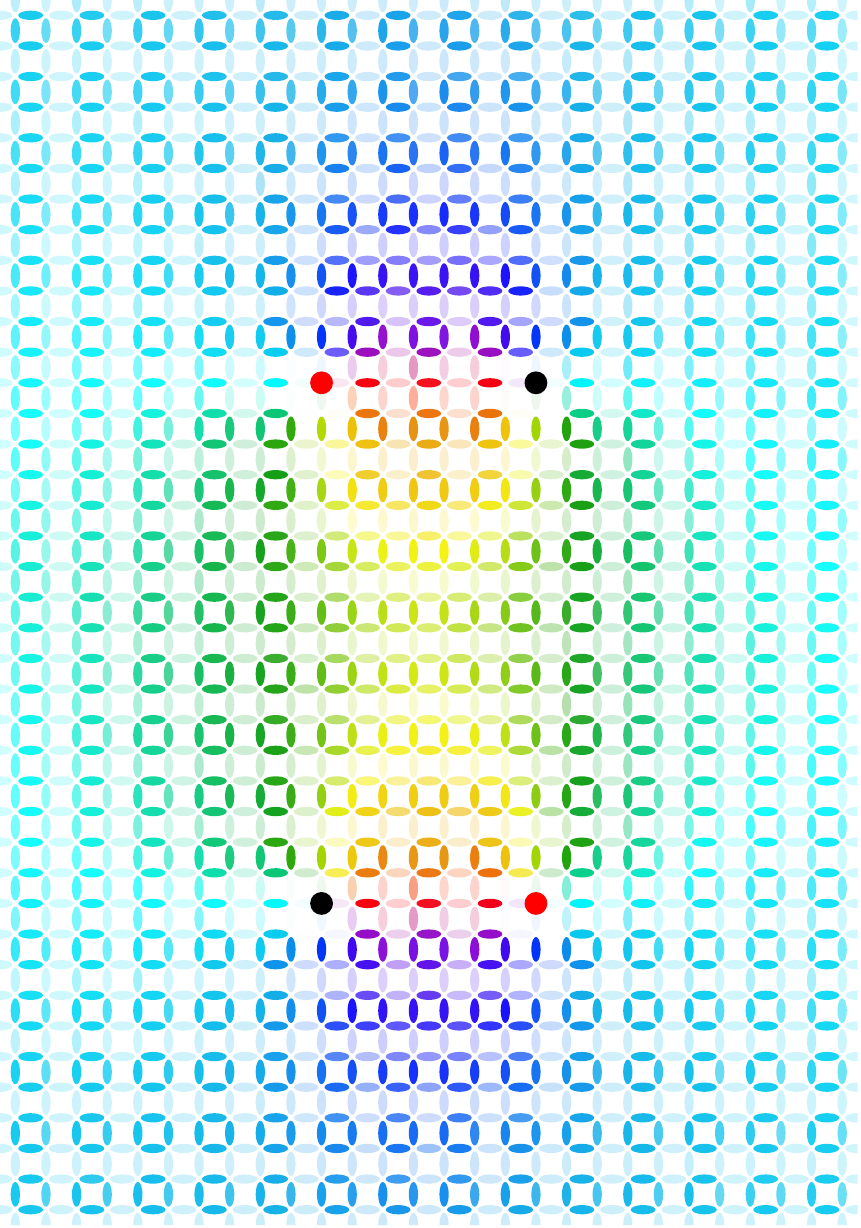}}
        \put(0.03,1.396){\makebox(0,0)[lt]{\smPanelTag{d}}}
    \end{picture}
    \caption{Real-space VBS patterns of the four-spinon string complex in the
    $L=48$ system at fixed $d_x=3$.  From left to right:
    (a)~$d_y=2$, (b)~$d_y=4$, (c)~$d_y=6$, and (d)~$d_y=8$.
    The spinon quadrupole is progressively elongated as $d_y$ increases, while
    the connected VBS domain-wall structure remains stable and shows no string
    breaking within the displayed range.  The larger separation and weaker
    overlap between neighboring strings reduce the inter-string repulsion
    relative to the spinon-dipole geometry.}
    \label{fig:vbs-quadruple-variation}
\end{figure}

Figure~\ref{fig:vbs-quadruple-variation} shows the corresponding evolution of
a four-spinon string complex.  We fix the horizontal pinned-spinon separation
at $d_x=3$ and increase the vertical separation from $d_y=2$ to $d_y=8$.
The quadrupolar VBS texture becomes progressively elongated, but its connected
string structure remains intact throughout this range.  Thus, unlike the
two-spinon dipole, no string breaking is observed for the quadruple geometry
within the present system size and boundary conditions.

The enhanced stability can be understood from the internal string geometry.
In the quadruple, each neighboring pair of spinons is connected by two VBS
domain-wall strings.  These strings are more widely separated than the
overlapping strings of the spinon dipole, reducing their overlap and the
associated inter-string repulsion.  The resulting lower interaction energy
favors an extended connected quadrupole even as the pinning geometry is
stretched.

\begin{figure}[htp]
    \centering
    \setlength{\unitlength}{0.24\linewidth}
    \begin{picture}(1,1)
        \put(0,0){\includegraphics[width=\unitlength]{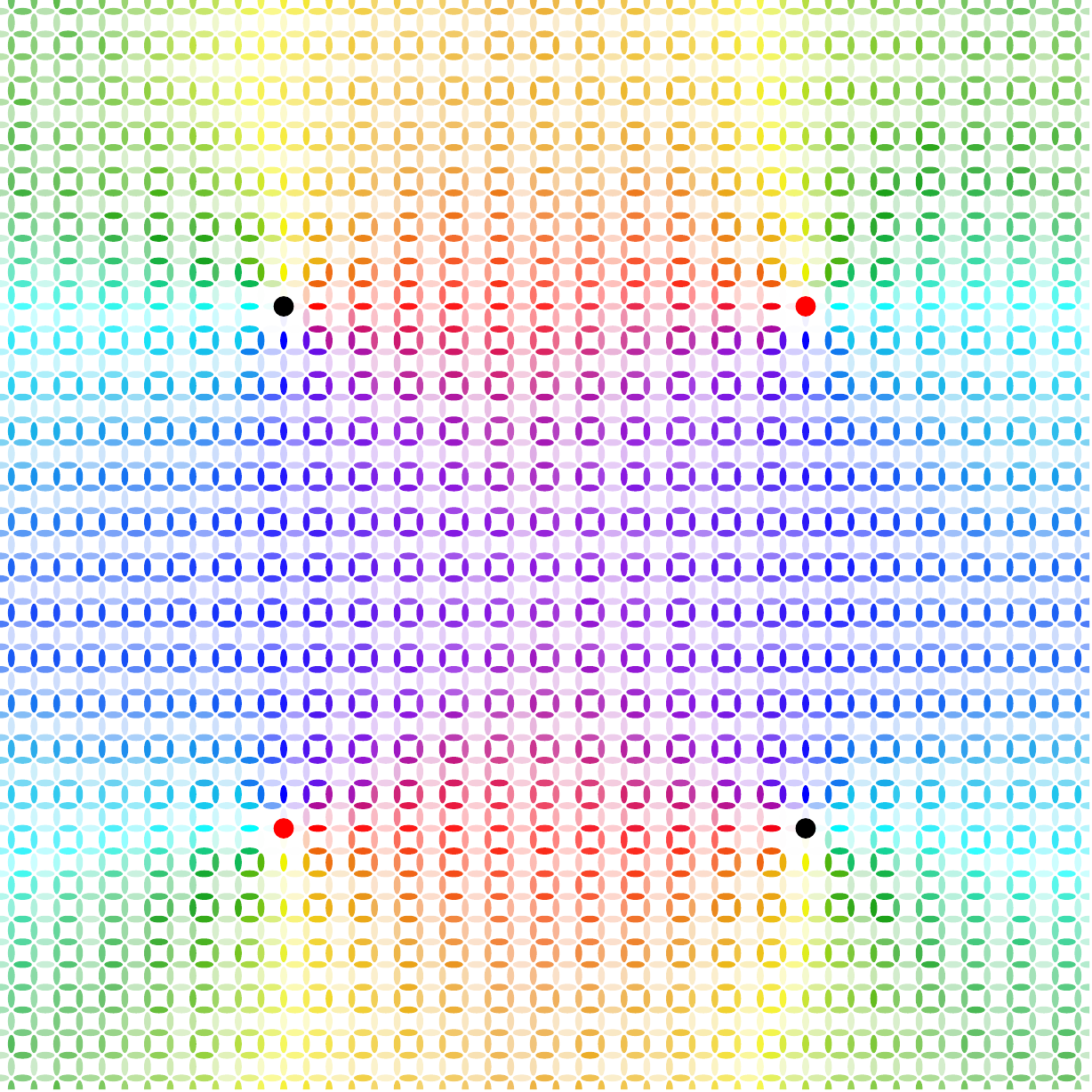}}
        \put(0.03,0.97){\makebox(0,0)[lt]{\smPanelTag{a}}}
    \end{picture}%
    \hspace{0.05\linewidth}%
    \begin{picture}(1,1)
        \put(0,0){\includegraphics[width=\unitlength]{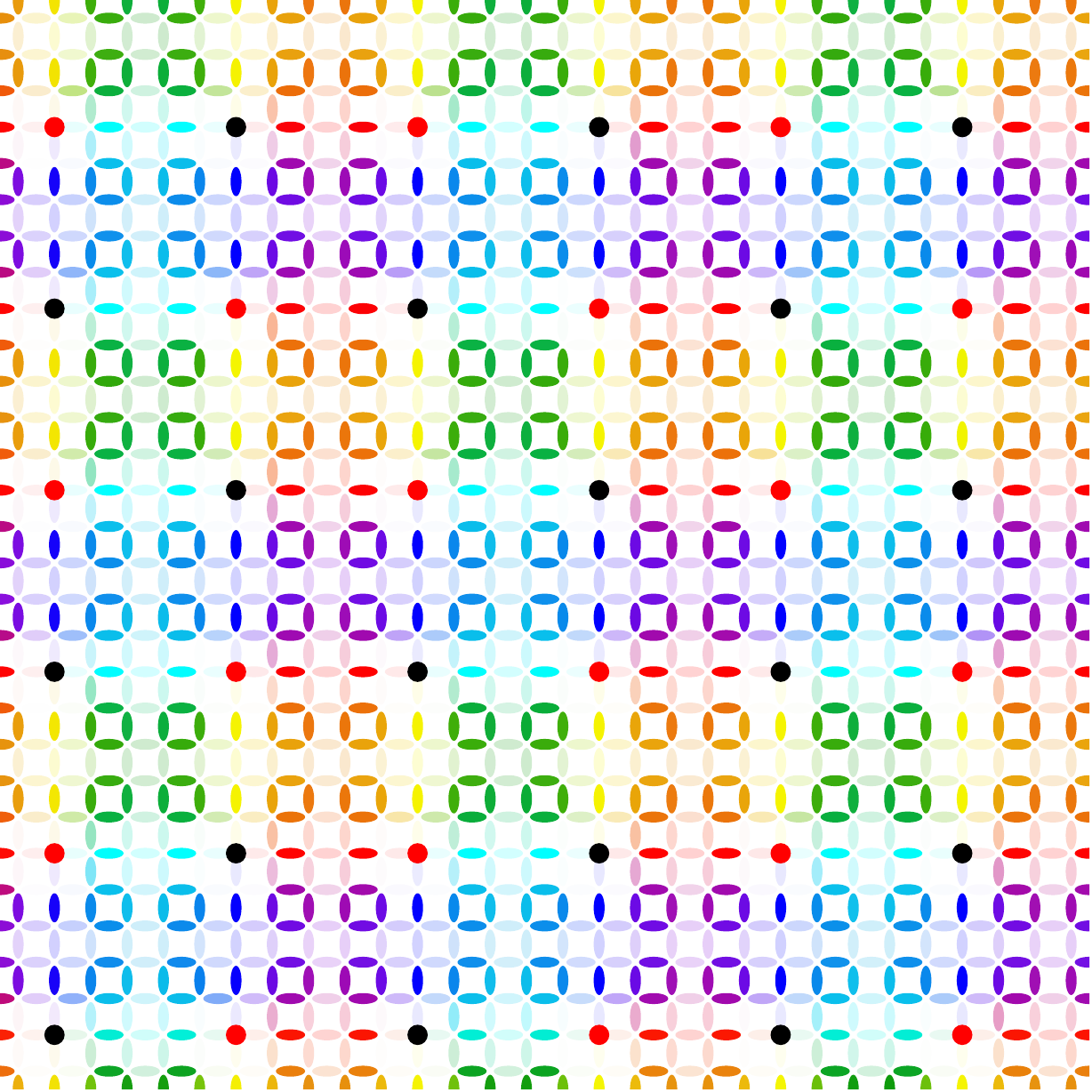}}
        \put(0.03,0.97){\makebox(0,0)[lt]{\smPanelTag{b}}}
    \end{picture}
    \caption{Representative tunable domain-wall networks generated by pinned
    spins.  (a) Four pinned-spin defects in an $L=48$ system with adjacent
    separations $23$ and $25$ under periodic boundary conditions.  Each
    neighboring pair is connected by a single long VBS domain-wall string,
    reducing inter-string overlap and repulsion and stabilizing strings over
    nearly the full system length.  (b) A dense pinned-spin array with spacing
    $d=5$, demonstrating that the lattice spacing and density of the
    domain-wall network can be adjusted through the pinning geometry.}
    \label{fig:dwn-variation}
\end{figure}

The same reduction of inter-string repulsion also stabilizes much more extended
structures.  Figure~\ref{fig:dwn-variation}(a) shows four pinned-spin defects in
an $L=48$ system, with adjacent separations of $23$ and $25$ along the periodic
direction.  Because of the periodic boundary condition, each neighboring pair
of pinned spinons is connected by a single VBS domain-wall string.  The strings
therefore have little mutual overlap and a correspondingly weak repulsion,
allowing them to remain stable over distances comparable to the system size.

This construction is not restricted to a sparse four-pin geometry.  The dense
array with pinning spacing $d=5$ in Fig.~\ref{fig:dwn-variation}(b) forms a
compact periodic domain-wall network.  Varying the positions and spacing of
the pinned spins therefore provides direct control over the lattice spacing
and density of the emergent network.

\section{VBS phase offset and domain-wall connectivity}
\label{app:vbs-phase-offset}

\begin{figure}[htp]
    \centering
    \includegraphics[width=0.3\linewidth]{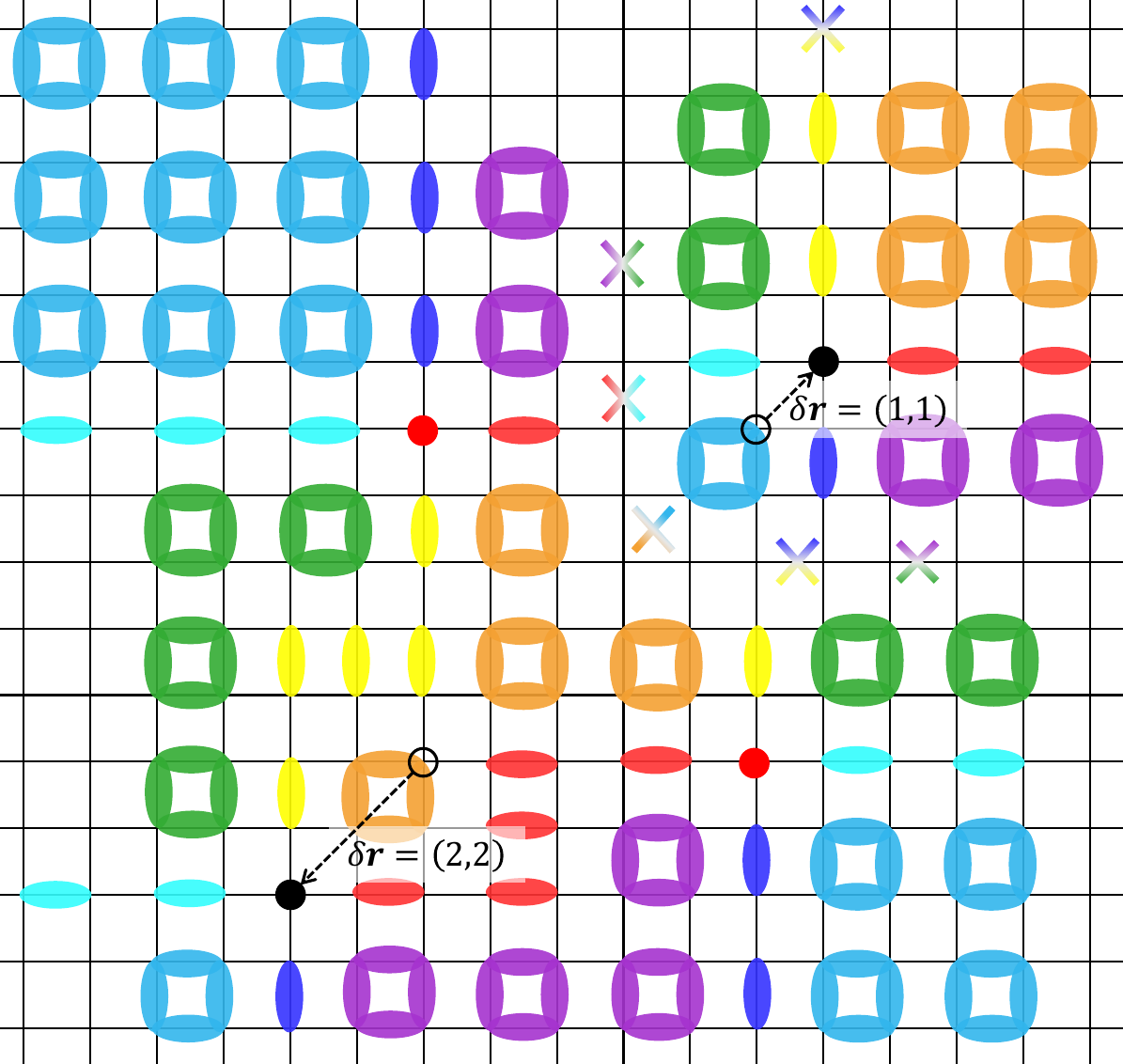}
    \caption{Schematic illustration of the lattice-scale VBS phase offset. A
    diagonal displacement $\delta\mathbf r=(1,1)$ preserves the spinon
    sublattice and vorticity but shifts the surrounding VBS phase by $\pi$,
    frustrating the original domain-wall matching and inducing local
    reconnection. For $\delta\mathbf r=(2,2)$, the accumulated $2\pi$ phase
    shift is trivial modulo $2\pi$, so the original connectivity is restored.}
    \label{fig:sm-vbs-phase-offset}
\end{figure}

In addition to its vorticity $q$ and spin projection $S^z$, a spinon inherits a
lattice-scale phase offset from the surrounding VBS pattern, as illustrated in
Fig.~\ref{fig:sm-vbs-phase-offset}. Translating a spinon by
$\delta\mathbf r=(1,1)$ preserves its A/B sublattice parity and therefore leaves
its VBS vorticity unchanged. Within our convention for the VBS order parameter,
however, the same translation shifts the surrounding VBS angle by $\pi$.
Although the displaced spinon thus carries the same continuum vortex charge,
its local VBS pattern is no longer phase matched to the original domain-wall
configuration. The existing strings must consequently rearrange or reconnect
near the displaced core.

By contrast, a displacement $\delta\mathbf r=(2,2)$ produces a total phase shift
of $2\pi$, which is equivalent to zero modulo $2\pi$. The original phase
matching, and hence the domain-wall connectivity, can then be restored. The
microscopic connectivity of spinon-bound domain walls is therefore constrained
not only by vorticity and spin quantum numbers, but also by a discrete
lattice-scale VBS phase offset that is absent from a continuum description based
only on vortex charge.

\section{VBS pattern in mismatched pinned configurations}
\begin{figure}[htp]
    \centering
    \setlength{\unitlength}{0.27\textwidth}
    \begin{picture}(1,0.989)
        \put(0,0){\includegraphics[width=\unitlength]{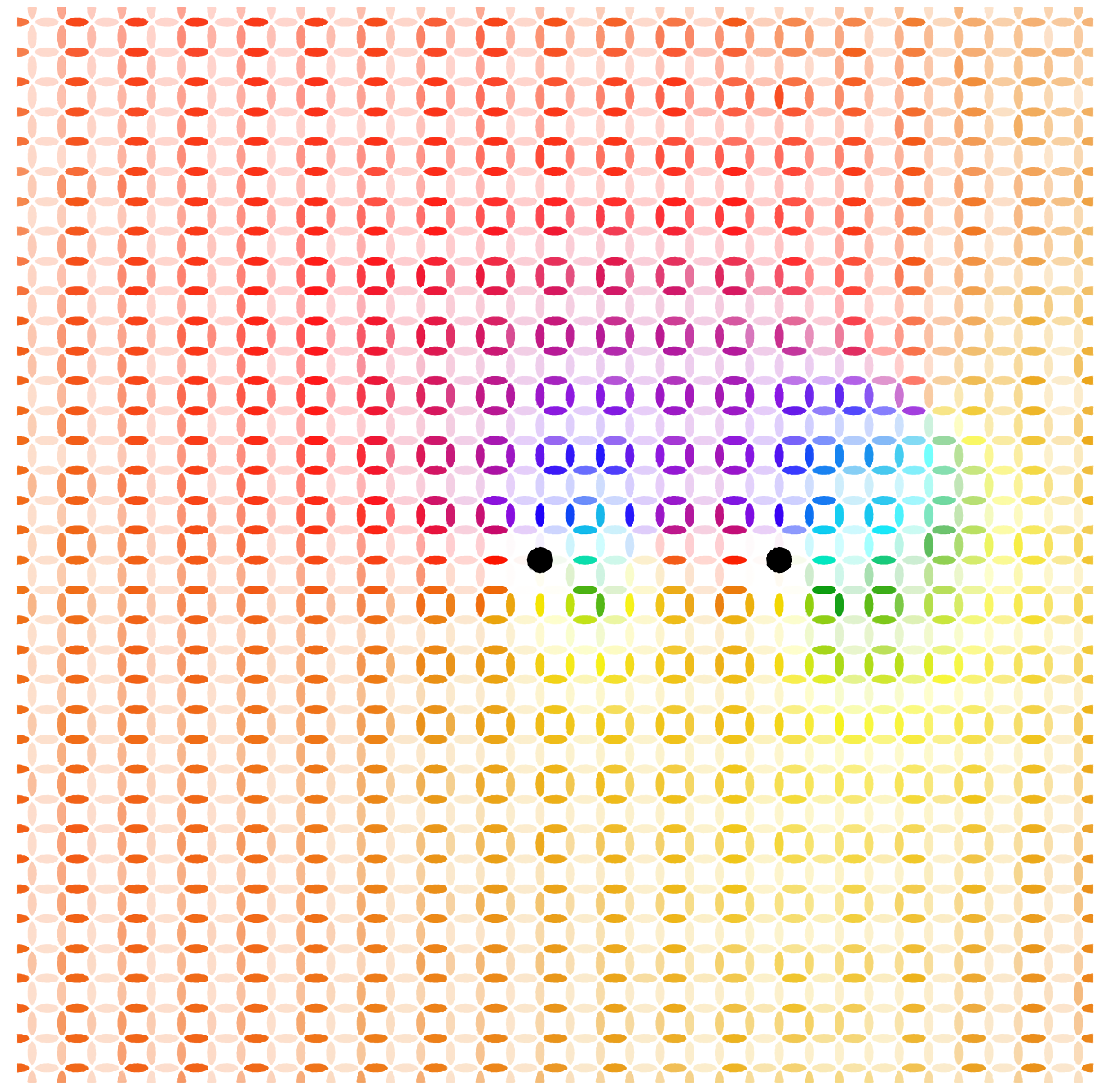}}
        \put(0.03,0.959){\makebox(0,0)[lt]{\smPanelTag{a}}}
    \end{picture}%
    \hspace{0.015\linewidth}%
    \setlength{\unitlength}{0.27\textwidth}
    \begin{picture}(1,1.005)
        \put(0,0){\includegraphics[width=\unitlength]{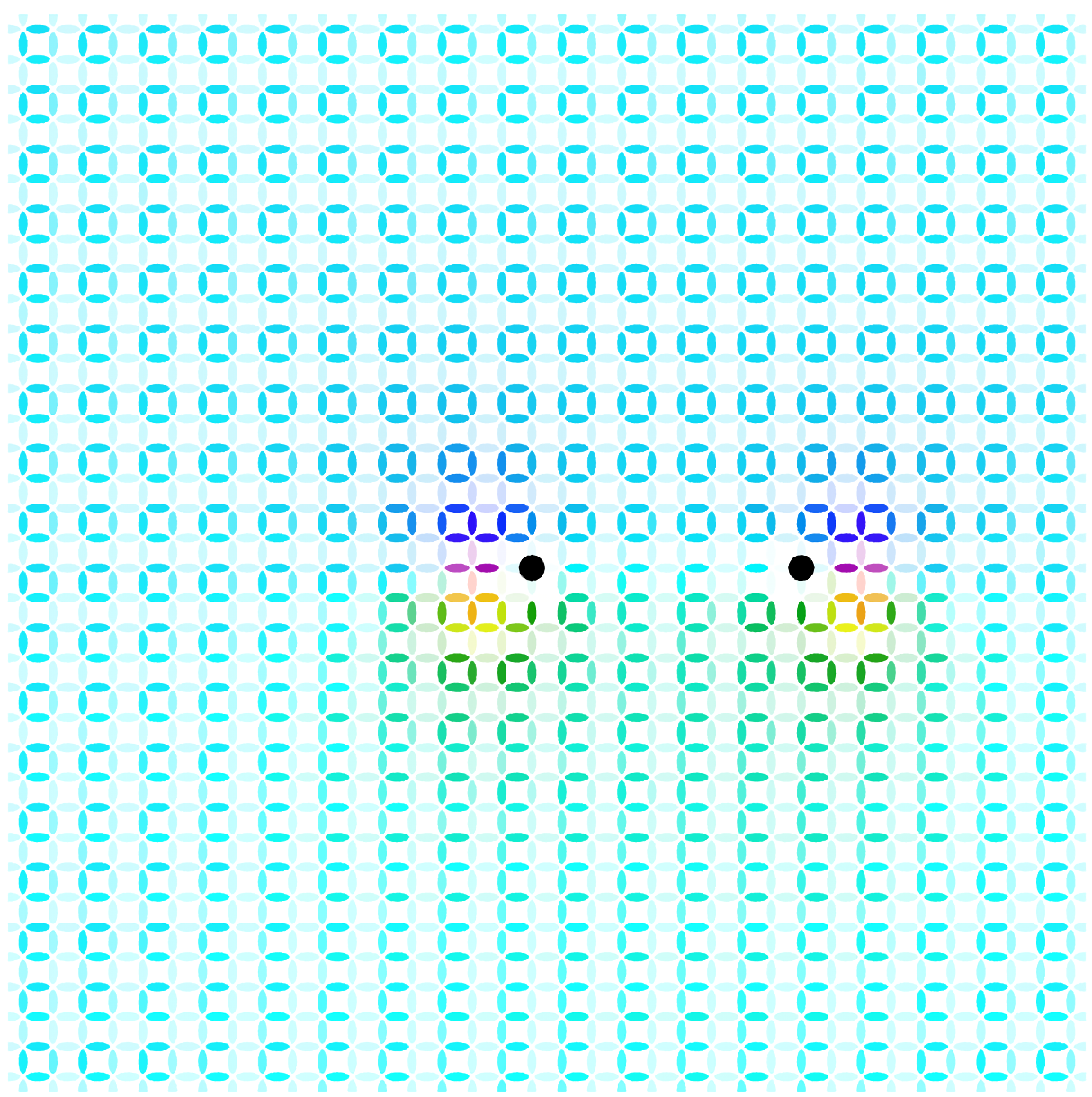}}
        \put(0.03,0.975){\makebox(0,0)[lt]{\smPanelTag{b}}}
    \end{picture}%
    \hspace{0.015\linewidth}%
    \setlength{\unitlength}{0.29\textwidth}
    \begin{picture}(1,0.942)
        \put(0,0){\includegraphics[width=\unitlength]{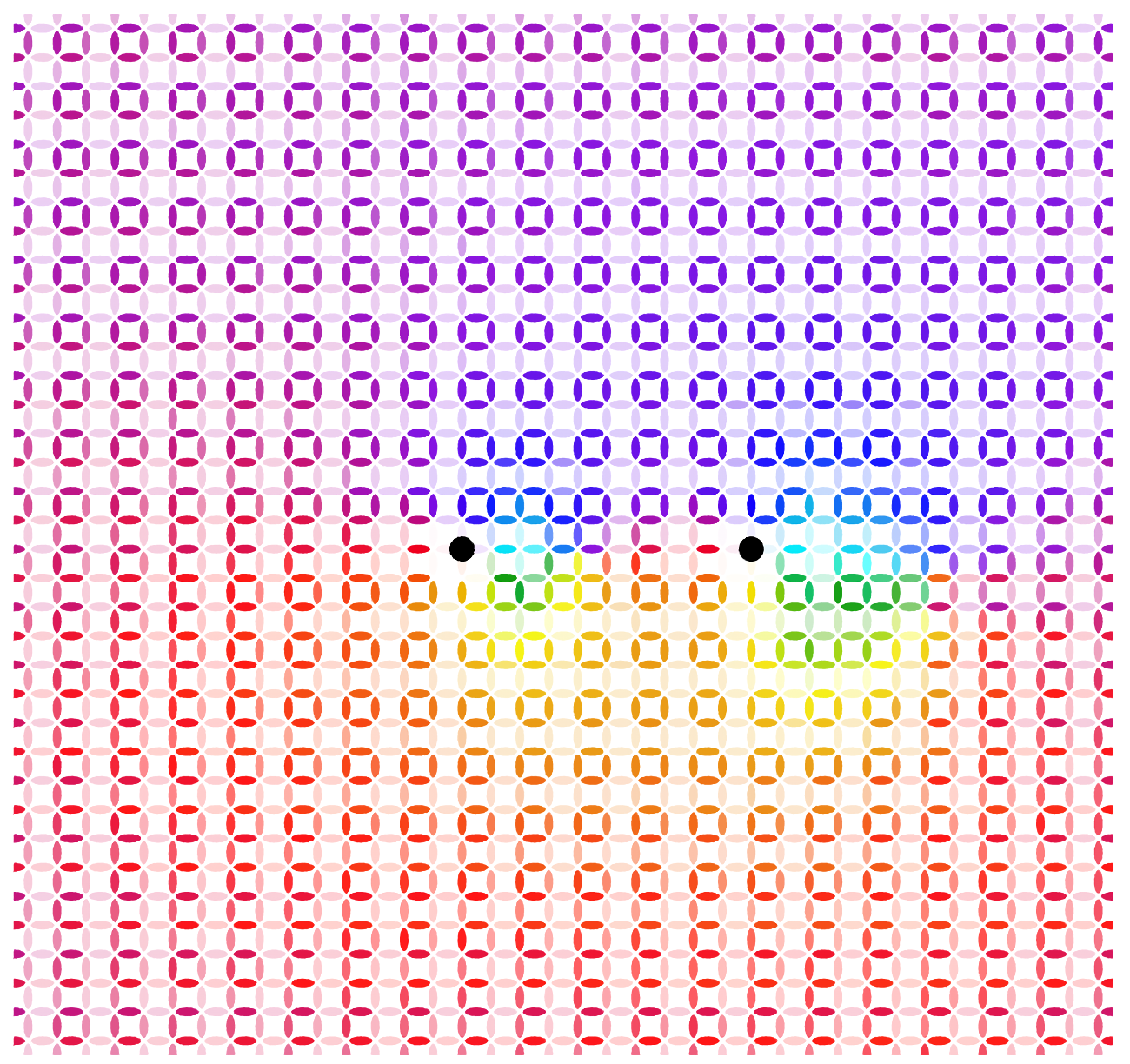}}
        \put(0.03,0.912){\makebox(0,0)[lt]{\smPanelTag{c}}}
    \end{picture}
    \caption{Two spins pinned at a distance $d$ in a $L=48$ system. The VBS pattern is shown for three different distances: (a) $d=8$ (same sublattice) and antiparallel pinned directions, (b) $d=9$ (different sublattice) and parallel pinned directions, and (c) $d=10$ (same sublattice) and parallel pinned directions. The two pinned spins are marked by black dots.}
    \label{fig:mismatched_pinned}
\end{figure}

Here we present the VBS pattern for three types of mismatched pinned configurations, as shown in Fig.~\ref{fig:mismatched_pinned}. If the gauge and spin constraints are not simultaneously satisfied by the two pinned spins, a pair of compensatory spinons are emerged to restore neutrality. The resulting VBS pattern is determined by the interplay of the spin and gauge interactions between the pinned spins and the compensatory spinons.
The resulting interaction has two channels. The spin channel is an RKKY-like magnetic exchange mediated by the spin susceptibility of the VBS background \cite{sachdev1999quantum,auerbach2012interacting,legg2019spin,ansari2024magnetic},
\begin{equation}
E_{\rm spin}(r)\sim J_{\rm eff}(r)S^z_1S^z_2,\qquad
J_{\rm eff}(r)\propto \chi_{zz}(r),
\end{equation}
which is short-ranged in the gapped VBS phase and becomes enhanced near the DQCP. This is only RKKY-like in structure, since the mediator is the insulating quantum magnet rather than a metallic Fermi sea. The gauge channel is associated with the vortex charge of the VBS angle. In the approximate $U(1)$ regime, it resembles the vortex interaction of an XY model, with like vortices repelling and vortex--antivortex pairs attracting. 
The spinons tend to bind into neutral pairs that carry no net spin or gauge charge but possess a finite separation, and therefore behave as effective dipoles. Interactions between such pairs are then analogous to dipole–dipole interactions in electrodynamics: they depend not only on the distance between pairs but also on their relative orientation and internal structure. As a result, the total energy of a configuration is governed by a complicated superposition of multi-dipole interactions, including both pairwise and collective contributions. The observed VBS texture is therefore the configuration that best satisfies the combined spin and gauge neutrality constraints while minimizing this intricate combination of interaction channels; a systematic treatment of these multi-dipole effects is beyond the scope of the present work and is left for future study.

\section{Real-space VBS pattern distribution}
\label{VBS-distribution}
The pinning geometry rearranges the locally selected VBS background.  To
characterize this response, we construct a spatial histogram of the
bond-resolved VBS order parameter.  For each bond \(b\), we write
\[
\Psi_{\rm VBS}(b)=T_x(b)+iT_y(b)=\rho_b e^{i\phi_b},
\]
where the bond-centered fields \(T_\mu(b)\) are defined by averaging the
site-centered fields at the two endpoints, as described in
Sec.~\ref{app:vbs-coarse-graining}.
After obtaining a coarse-grained real-space texture, we bin all bonds in the
polar plane \((\rho_b,\phi_b)\).  The value in each bin is the fraction of
bonds with the corresponding local VBS order, normalized by the total number
of bonds in the sampled region.  This histogram therefore describes the
spatial distribution within a single coarse-grained texture rather than
statistical fluctuations among independent QMC samples.  A concentrated
bright region identifies the dominant local VBS order selected by the imposed
pinning geometry.

\begin{figure}[htp]
    \centering
    \includegraphics[width=0.7\linewidth]{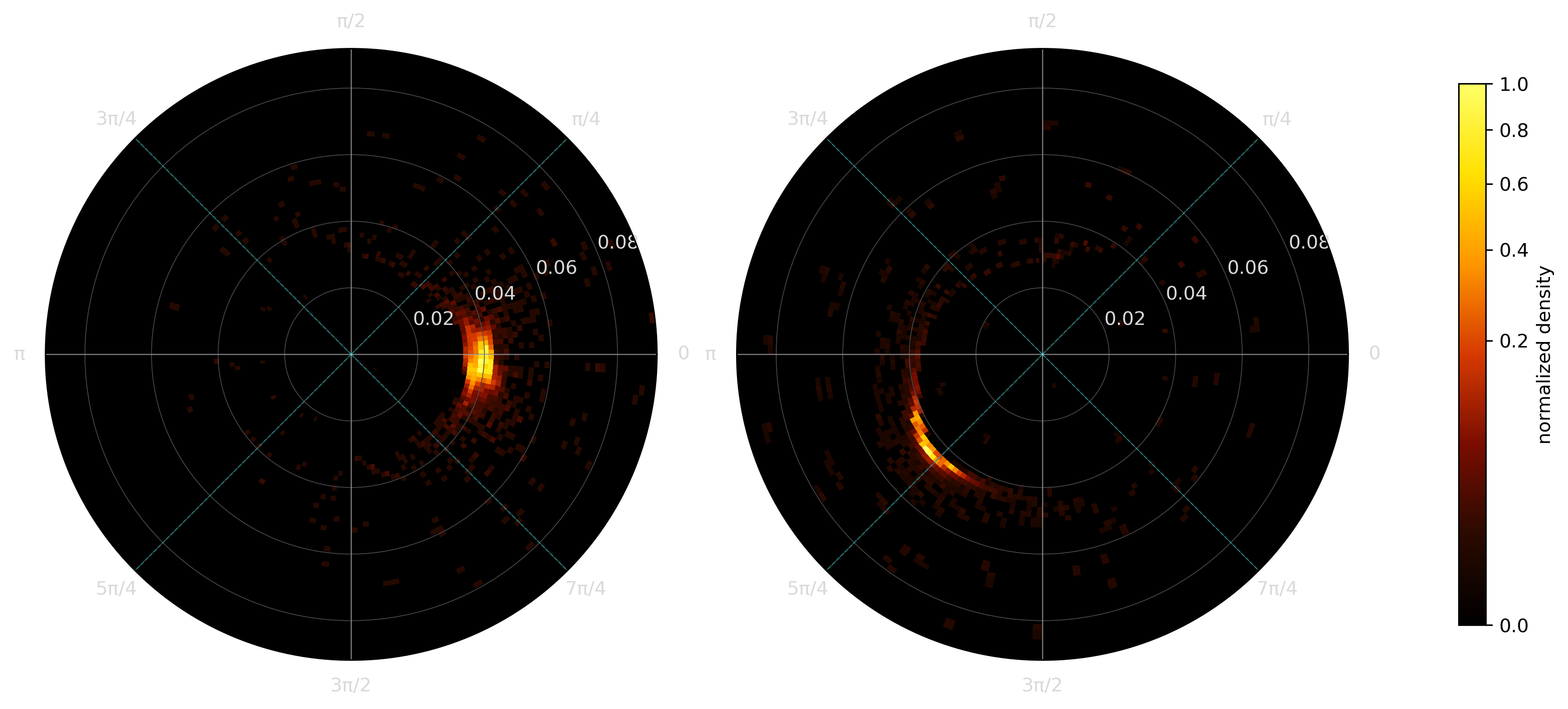}
    \caption{Spatial histogram of the local VBS order parameter distribution in the coarse-grained real-space texture. The color represents the normalized fraction of bonds. We compare (left) the unbroken-string two-pin dipole texture in Fig.~1(c) of the main text with (right) the four-pin quadrupole texture shown in Fig.~2(b) of the main text.}
    \label{fig:vbs-heatmap}
\end{figure}

Figure~\ref{fig:vbs-heatmap} compares the resulting distributions.  The left
panel is obtained from the unbroken configuration shown in Fig.~1(c) of the
main text, while the right panel is obtained from the four-pin configuration
in Fig.~2(b) of the main text.  The two-pin unbroken texture has most of its
spatial weight near a columnar-VBS angle, whereas the four-pin quadrupole
texture has most of its weight near a plaquette-VBS angle.  Thus, the four-pin
geometry locally selects a plaquette-like VBS background inside the
quadrupolar texture, without implying a bulk phase transition.

\section{\(Z_4\)--\(U(1)\) crossover}

\begin{figure}[htp]
    \centering
    \includegraphics[width=0.45\linewidth]{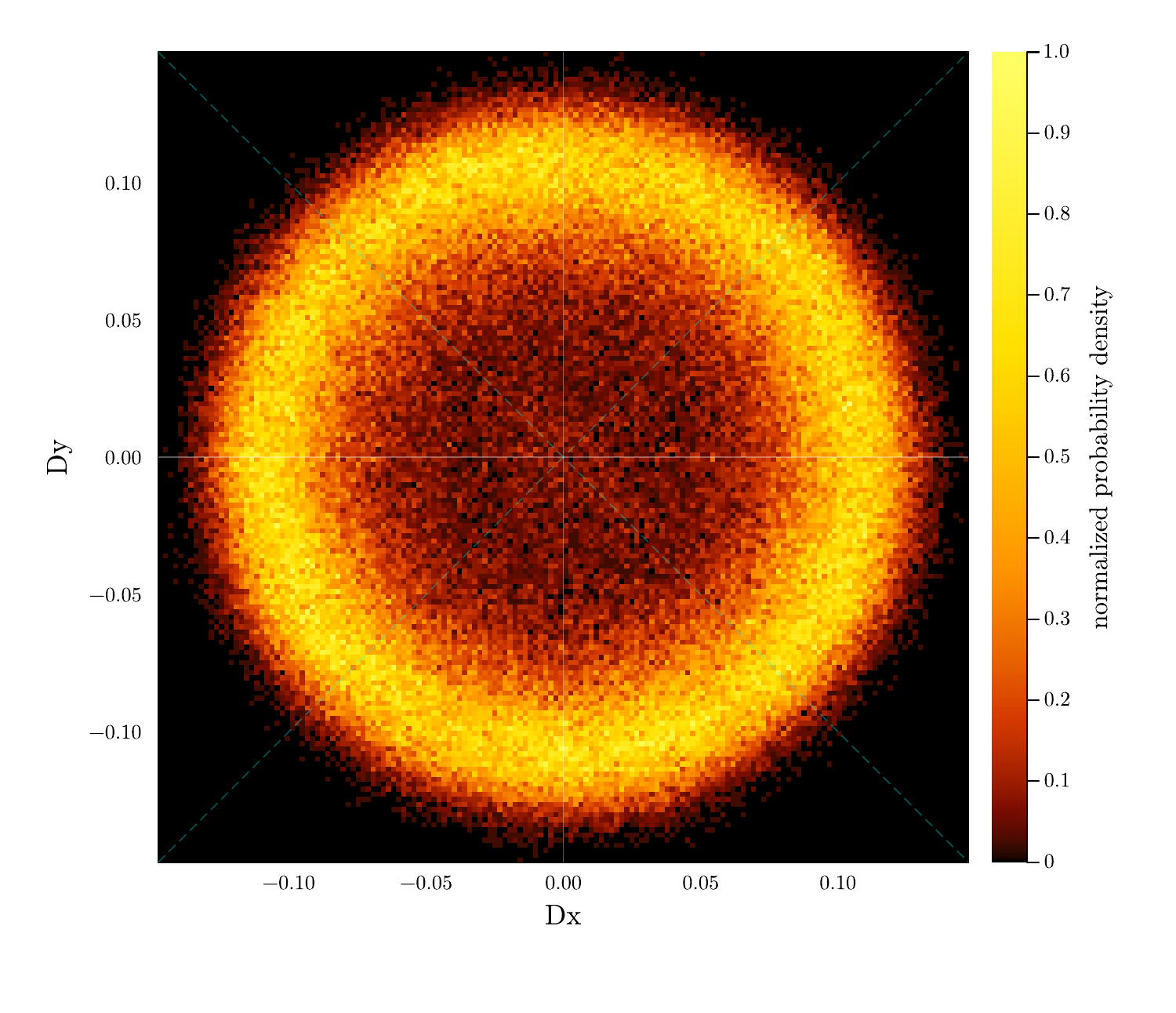}
    \caption{Probability density of the global VBS order parameter
    \(\mathbf D=(D_x,D_y)\) for the \(L=48\) pure-\(Q_2\) system.  The density
    is normalized by its maximum, and black denotes zero sampled density.  The
    nearly circular annulus indicates that the fourfold anisotropy is weak on
    this length scale.}
    \label{fig:global-vbs-distribution}
\end{figure}

Figure~\ref{fig:global-vbs-distribution} shows the distribution of the global
VBS order parameter \(\mathbf D=(D_x,D_y)\) for the \(L=48\) pure-\(Q_2\)
system.  The distribution forms a nearly circular annulus in the
\(D_x\)--\(D_y\) plane, without pronounced peaks along the four columnar VBS
directions.

We quantify the fourfold anisotropy by
\begin{equation}
W_4\equiv
\left\langle\cos\!\left(4\phi_{\rm VBS}\right)\right\rangle .
\end{equation}
Here \(W_4=1\) corresponds to weight concentrated along the four columnar
directions
\(\phi_{\rm VBS}=0,\pi/2,\pi,3\pi/2\), whereas \(W_4=-1\) corresponds to the
four diagonal or plaquette directions
\(\phi_{\rm VBS}=\pi/4,3\pi/4,5\pi/4,7\pi/4\).  A value \(W_4=0\) indicates no
detectable net fourfold angular anisotropy, as expected for a uniform
\(U(1)\) distribution.

For the present data, \(W_4=0.0083(75)=0.0083\pm0.0075\), where the uncertainty
is obtained from the scatter among eight independent QMC chains.  This value
is statistically consistent with zero. 
This, together with the nearly circular
distribution in Fig.~\ref{fig:global-vbs-distribution}, places the \(L=48\)
system size to be in an effectively $U(1)$ symmetric regime~\cite{lou2009z,shao2016quantum}, which should ultimately become $Z_4$ in the thermodynamic limit. 

\end{document}